\documentclass[12pt]{iopart}
\usepackage[latin1]{inputenc}
\usepackage{amsfonts}
\usepackage{amssymb}
\usepackage{graphicx}
\newtheorem{thm}{Theorem}

\begin{document}
\title{A minimal model for congestion phenomena on complex networks}
\author{Daniele De Martino$^{1}$, Luca Dall'Asta$^{2}$, Ginestra Bianconi$^{2}$ and Matteo Marsili$^{2}$}

\address{$^1$ International School for Advanced Studies and INFN, via Beirut 2-4, 34014 Trieste (Italy)}
\address{$^2$ The Abdus Salam International Centre for Theoretical Physics, Strada Costiera 14,
34014 Trieste (Italy)}

\begin{abstract}
We study a minimal model of traffic flows in complex networks, 
simple enough to get analytical results, but with a very rich
phenomenology, presenting continuous, discontinuous as well as hybrid phase transitions
between a free-flow phase and a congested phase, critical points and 
different scaling behaviors in the system size. 
It consists of random walkers on a queueing 
network with one-range repulsion, where particles can be destroyed only if they can move.
We focus on the dependence on the topology as well as on the level of traffic control. 
We are able  to obtain transition curves and phase diagrams at analytical level for the
ensemble of uncorrelated networks and numerically for single instances.
We find that traffic control improves global performance, enlarging the free-flow region 
in parameter space only in heterogeneous networks. 
Traffic control introduces non-linear effects and, beyond a critical
strength, may trigger the appearance of a congested phase in a
discontinuous manner.
The model also reproduces the cross-over in the
scaling of traffic fluctuations empirically observed in the Internet, and
moreover, a conserved version can reproduce 
qualitatively some stylized facts of traffic in transportation networks.
\end{abstract}

\section{Introduction}

The aim of applying methods of statistical physics to the complex behavior of
traffic in large networked infrastructures is to identify the most important statistical 
regularities and explain the origin of the collective phenomena observed in real systems, such as the Internet.
As many other complex systems, infrastructure networks can be described and studied at different levels of 
detail, therefore one of the main issues consists in recognizing which ingredients are relevant at a given scale, neglecting all
redundant information. 

In this perspective, many collective phenomena occurring on complex networks can be seen 
as emerging from simple microscopic rules, such as the dynamics of diffusing and interacting particles on graphs \cite{PV04} .
Examples of real phenomena that can be explained using simple dynamical processes range from avalanches of failures in 
power-grids \cite{M04}, to credit contagion in networks of firms \cite{HK08} and the diffusion of e-mail viruses and spams \cite{PV01}.
The complex phenomena related to traffic, both in transportation \cite{PCL06,DMM04}.
and  communication networks \cite{ADG01,TT,EGM05,TTR04,GDVCA02} have been similarly
tackled using simplified descriptions aimed to characterize their main statistical properties using concepts and 
methods that are typical of non-equilibrium statistical mechanics.
The approach of statistical mechanics allows to disentangle the complex collective behavior of these systems, providing tools to prevent failures and  to improve their performances.

However, the fact of dealing with simple interacting particles systems is sometimes perceived as a limit to the richness and variety of the observed phenomena. 
An emblematic counterexample is provided by the discovery that particles condensation can occur in one of the simplest classes of non-equilibrium processes on graphs, the Zero-Range Processes \cite{EH05}. The possibility to study these processes analytically, instead of resorting to numerical simulations, is of great help to elucidate the properties of condensation phenomena in more realistic processes of mass-transfer and traffic on networks.  

In the same spirit we propose in this work a simple model of traffic on network in the form of a system of particles that are free to hop randomly between nodes but are subject to the constraint of forming queues at the nodes. This model was recently proposed by the authors as a paradigm to study congestion phenomena on networks \cite{DDBM09}. 
Hence, our focus is here congestion in general, as a collective phenomenon arising in particles systems in particular dynamical regimes.
The mechanisms responsible of the phase transition from a free-flow to a congested stationary state are discussed in detail, explaining under which conditions the observed transition is continuous or discontinuous.  
Even though the dynamical process is presented in the very general and idealized framework of interacting particles systems, our results have an immediate application in the study of several traffic systems, from the Internet to road networks. For instance, in the light of our analysis the validity of traffic control strategies in order to prevent congestion is questioned. 

The paper is organized in the following way. In Section \ref{sec-model} we present our model of dynamical process on graph, discussing its relation with other classes of well-known interacting particles systems. A brief account of the phenomenology of the model is given in
Section \ref{sec-model2}. Section \ref{sec-theory} is instead devoted to study the stationary state behavior by means of analytical approaches. We provide an iterative method to characterize the stationary state on every given graph (Section \ref{sec-theory1}) and a mean-field analysis at the level of random graphs ensembles (Section \ref{sec-theory2}). Finally, in Section \ref{sec-theory3}, we discuss the relation with condensation phenomena observed in other particles systems. 
The applications to Internet and vehicular traffic are discussed in Section \ref{sec-applications}.  We conclude and indicate some possible future development in Section \ref{sec-conclusions}.  \ref{app1} and \ref{app2} are devoted to discuss under which conditions a product-measure stationary probability distribution  exists. \ref{app3} discusses the case of small queueing capacity, that does not change considerably from the large capacity limit considered in the main text.

\section{The model}
\label{sec-model}

\subsection{Definition of the model}
\label{sec-model1}
Let us consider a network of $N$ nodes and let $v(i)$ be the set of neighbors of node $i$. 
We describe particles dynamics as a continuous time stochastic process, in which particles are generated at each node $i$
with a rate $p_i$. Each node is endowed with a first-in first-out queue, in which arriving particles are stored. 
Let $n_i$ be the number of particles in the queue of node $i$. If $n_i>0$, we
assume the following probabilistic hop rule: the topmost particle leaves the node at a rate $r_i$ and jumps
in the queue of a randomly chosen neighbor $j \in v(i)$. 
With probability $\eta(n_j)$ the particle is rejected by the arrival node and remains on the departure one. 
Otherwise, particles are either destroyed during the hopping,  with a probability $\mu_j$, or, with probability $1-\mu_j$, enter the queue on node $j$. \\
Our model takes an Eulerian perspective, which focuses on the statistics of the length of the queues $\{ n_i \}$, rather than on the trajectories of particles. This perspective allows us to disregard the fate of individual particles, by replacing the processes by which they are generated and routed to their destination with probabilistic events.

The long-time behavior of the system is determined by the relation between creation and absorption
rates. Indeed, if creation rates are much larger than the absorption rates, the queues receive more particles of what they can dispose of
and the network is rapidly overloaded by particles. 
Such a dynamics leads to the onset of a {\em congested state}, where queues grow indefinitely. 
As the external drive imposed by the creation rates does not change in time,  
 the system eventually enters a {\em non-equilibrium stationary state} in which queues grow with constant velocity. 
In the non-equilibrium stationary state, a good observable is provided by   
the rate of growth of the total number of particles in the system \cite{ADG01}, 
\begin{equation}\label{op}
\rho = \lim_{t \to \infty} \frac{\mathcal{N}(t+\tau)-\mathcal{N}(t)}{\tau N p} 
\end{equation}
where $\mathcal{N}(t)=\sum_{i}n_i(t)$ is the total number of particles in the system at time $t$,
$p=N^{-1}\sum_i p_i$ is the average creation rate and $\tau$ is the observation time. 
Note that a local order parameter, replacing $\mathcal{N}(t)$ by $n_i(t)$ and $p$ by $p_i$, can be defined in the same way.
A node is thus {\em congested} if, in the stationary state, its average number of particles increases with time
($\rho_i \equiv \langle\dot n_i\rangle/p_i>0$).

We have already studied this model in the context of packet-transfer based communication networks \cite{DDBM09}, showing the existence of a phase transition from a {\em free-flow} regime to a {\em congested} state. As we will see later, the character of the phase transition depends on the parameters of the model and on the topology of the underlying network.

A conserved version of the model, without particles' creation and absorption,  can be defined as well:  
$\mathcal{N}$ particles are initially distributed randomly on the nodes, then the system is left 
evolve towards some stationary state in which the distribution of the lengths of the queues does not change anymore.
The distribution of particles on the nodes depends on the density of particles $\rho = \mathcal{N}/N$, on the local hop rates $\{ r_i \}$ and on the topology of the underlying network. This system presents a {\em condensation } phenomenon, in which a finite fraction of particles tends to occupy a single node or few classes of nodes.  

The model defined here is reminiscent of several other particles systems that have been recently studied by physicists and mathematicians, such as the Zero-Range Processes (ZRPs), the Misanthrope Processes (MPs) \cite{EH05}, and queueing networks \cite{BGDT06}.
In ZRPs, the hop rate depends only on the number of particles in the departure site, whereas in a MP it depends on both departure and arrival sites.
The steady-state distributions  $\mathcal{P}(n_1, \dots, n_N)$ of ZRPs admit a factorized form, thus these models are exactly solvable on every graph. 
For their simplicity, ZRPs are used as a theoretical test-ground for the study of the statistical properties of non-equilibrium systems.
Including in the hop rate a dependence on the arrival node, the process becomes a MP, that is also exactly solvable with factorized steady-state under very mild conditions (see e.g. Ref. \cite{EH05}).
The conserved version of our model belongs to this class of processes, with hop rates between connected nodes $i$ and  $j$,  
$u_{ij}(n_i,n_j) = r_i[1-\eta(n_j)]/k_i$.
Hence, the probability distribution $\mathcal{P}(n_1, \dots, n_N)$ defining the state of the system factorizes in the steady-state in a product measure over single-site distributions $\mathcal{P}_i(n_i)$ (see \ref{app1}). 

One would be tempted to extend the factorized form to the non-conserved model as well. In the \ref{app1}, we show under which conditions on the transition rates factorization is exact, while \ref{app2} reports some results on the general non-conserved model with particle rejection, from which it is possible to derive an approximated mean-field approach.

The choice of the functional form for the rejection probabilities $\eta(n_j)$ is motivated by the application of the model to several problems from the Internet's dynamics to vehicular traffic (see Section \ref{sec-applications}). 
In the following, $\eta(n_j)=\bar\eta \theta(n_j-n^*)$, that is node $j$ refuses particles with a probability $\bar\eta$ when it is already occupied by $n \geq n^*$ particles. 
An advantage of using this expression is that as long as the nodes have less than a threshold value $n^{*}$ of particles, the model
behaves as a ''grand-canonical'' ZRP, so factorization effectively holds. Hence, we expect that in the uncongested state, the product measure 
$\mathcal{P}(n_{1}, \dots, n_{N}) = \prod_{i}
\mathcal{P}_{i}(n_{i})$
 gives a very good approximation to describe the stationary asymptotic regime of the dynamics. This is also confirmed by the absence of correlations between  queues  of different nodes observed in the uncongested phase (see \ref{app2}).

\subsection{Phenomenology by simulations}
\label{sec-model2}

In the previous section we have announced the existence of a phase transition between a regime in which particles can freely move in the network and a congested one. This can be easily observed simulating numerically the model on any network and varying the value of the parameters. 
Let us consider for the moment the non-conserved model with homogeneous parameters $\mu_i=\mu$, $p_i=p$ and $r_i=1$ for all $i$. The simulations are performed on a random regular graph (of fixed degree $K=4$) and on an uncorrelated random graph with degree distribution $P(k) \propto k^{-\gamma}$ with $\gamma = 3$.

\begin{figure}[h]
\begin{center}
\includegraphics*[width=0.5\textwidth]{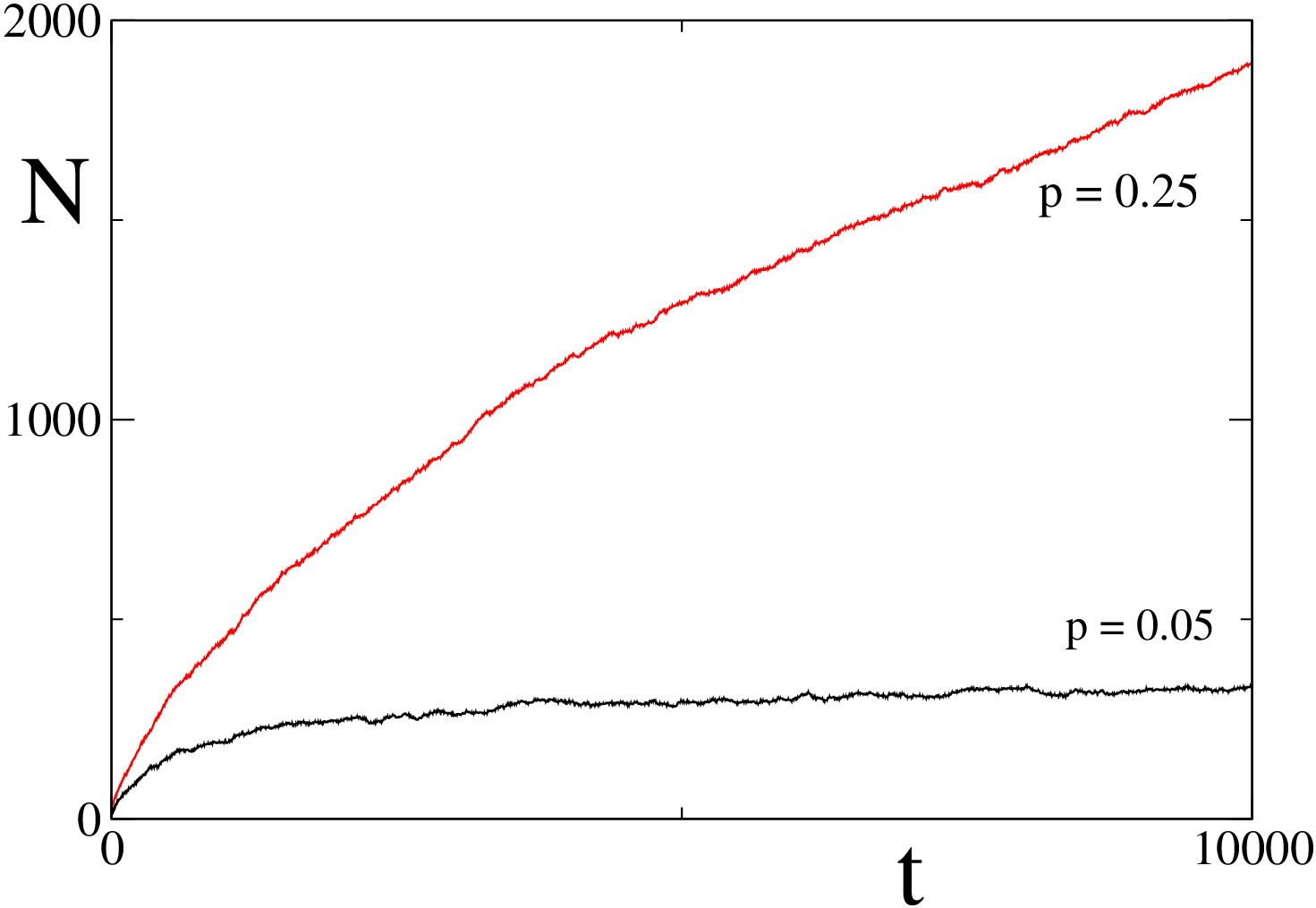}
\caption{Number of packets as a function of time for an homogeneous network of $1000$ nodes, degree $K=4$, without routing procol ($\eta=0$)
$\mu=0.2$, in the free ($p=0.05$) and congested phase ($p=0.25$)}
\label{phases}
\end{center}
\end{figure}

Figure \ref{phases}  displays two typical time series of the total number of packets $\mathcal{N}(t)$ in the free-flow  and congested  phases, obtained varying the creation/absorption ratio $p/\mu$. In the free-flow phase, the number of particles fluctuates 
around a stationary value that is much smaller than the network's size $N$, i.e. most of the nodes are empty. In the congested phase, instead, the number of particles waiting in the queues constantly increases in time, with the lack of stationarity in the number of particles.
In this example the congested phase appears around $p_{c} = \mu$, that seems trivial if we notice that a single-queue server with constant arrival rate $p$ and departure rate $\mu$ becomes overloaded exactly at $p/\mu =1$.  In fact,  the threshold $p_c = \mu$ is correct only on homogeneous networks (random regular graphs, regular lattices, etc.), not in the presence of large degree fluctuations.  

\begin{figure}
\begin{center}
\includegraphics*[width=0.35\textwidth]{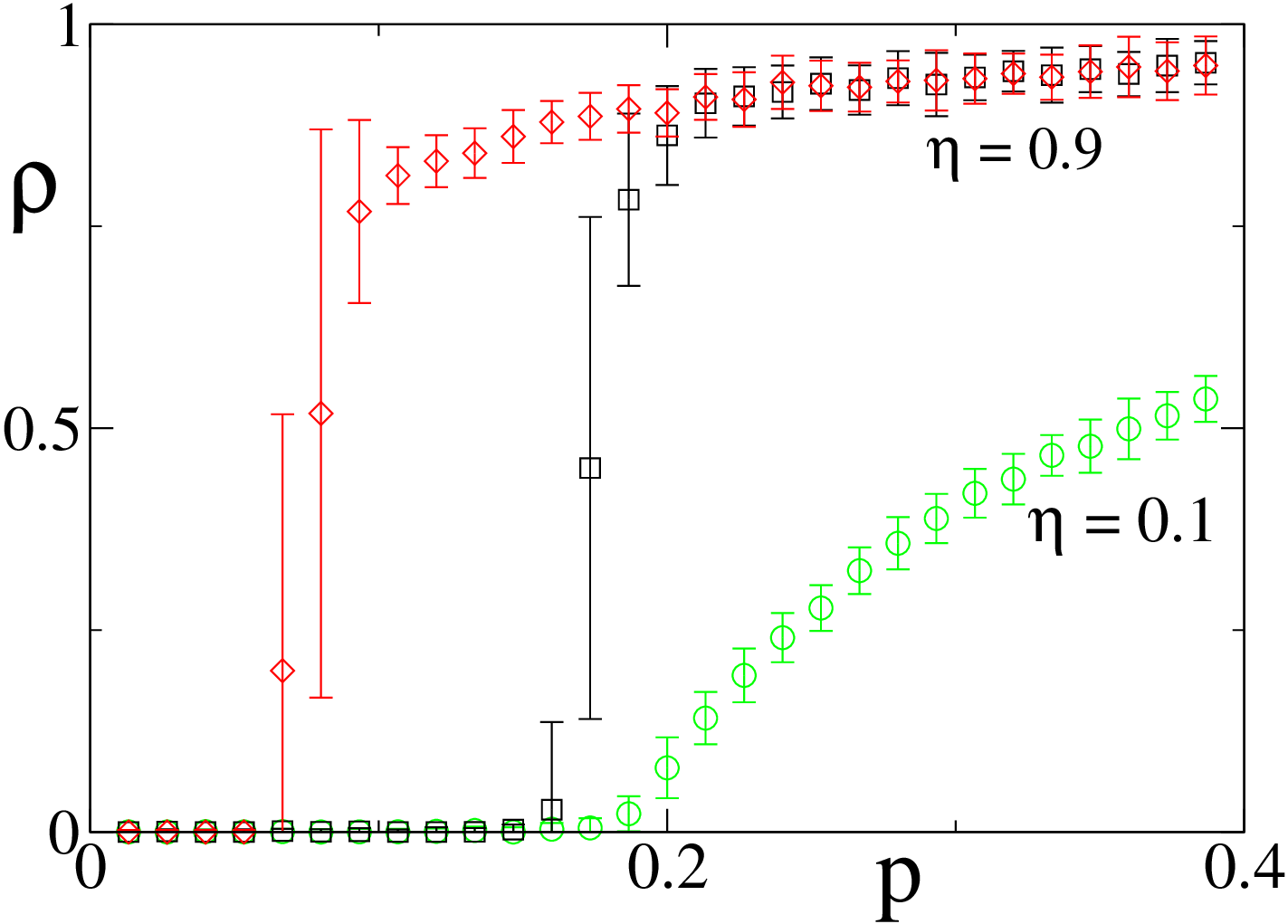}
\includegraphics*[width=0.35\textwidth]{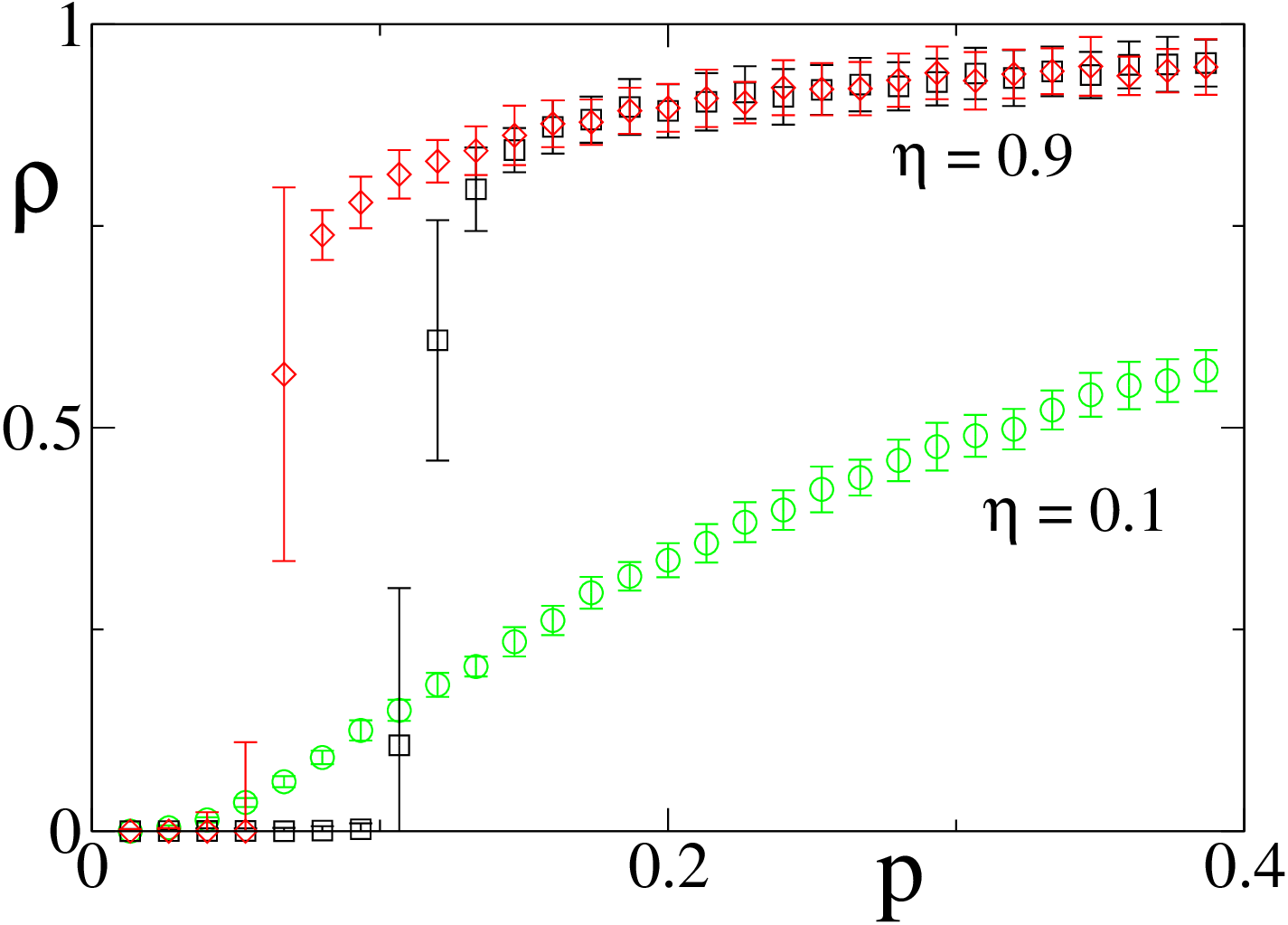}
\caption{Left: Transition curves $\rho(p)$ for a random regular graph of size $N=10000$, 
$\mu=0.2$, $\bar{\eta}=0.1$, $\bar{\eta}=0.9$. \\ Right: 
Transition curves $\rho(p)$ for an uncorrelated scale free graph with $\gamma=3$, $k_{min}=2$
of $N=3000$ nodes, $\mu=0.2$, $\bar{\eta}=0.1$, $\bar{\eta}=0.9$ for $n^*=10$.  
For $\eta=0.9$ the system shows {\em hysteresis} in both the homogeneous and heterogeneous case. 
}
\label{rhop}
\end{center}
\end{figure}

The difference between homogeneous and heterogeneous networks, as well as the role played by particles rejection, becomes evident looking at the behavior of the congestion parameter $\rho(p)$.  
In Fig. \ref{rhop} (left) we report $\rho(p)$ for a random regular graph in two different situations: with small rejection ($\bar{\eta}=0.1$, $n^*=10$) and with  strong rejection ($\bar{\eta}=0.9$, $n^* = 10$). The same plots for a scale-free network with $\gamma = 3$ are reported in Fig. \ref{rhop} (right). Let us focus on the curves obtained with small particles rejection. 
The homogeneous network becomes congested with a continuous phase transition taking place at $p_{c} \simeq \mu$, whereas for the heterogeneous one the congested state appears much earlier at $p \ll \mu$.  
In both cases, the effect of the increasing of $\bar{\eta}$ is that of changing the nature of the phase transition from continuous to discontinuous
with hysteresis. 
The continuous and discontinuous transitions (coming from lower $p$) are located at the same point $p_c \simeq \mu$ on homogeneous networks; on heterogeneous networks, instead, the discontinuous critical point is shifted towards larger values of the creation rate (with respect to the continuous critical point).

The theoretical understanding of these congestion phenomena depending on the hop rules and the underlying networks, as well as their potential application in the study of real traffic problems are the subjects of the next sections. 

\section{Analytical approach to the stationary state}
\label{sec-theory}
\subsection{Iterative equations on single graphs}
\label{sec-theory1}
In the conserved model, the single graph analysis is simple because factorization is exact, and the result on any given graph is exposed in \ref{app1}. The situation is different when particles are created and destroyed. 
Nevertheless, correlations between $n_i$'s on different nodes are hardly detectable in numerical simulations, as shown in \ref{app2}. This, and the fact that the interactions are local, makes it reasonable to  work within an factorized approximation, i.e. $\mathcal{P}(n_{1}, \dots, n_{N}) \simeq \prod_{i} \mathcal{P}_{i}(n_{i})$, and derive some local-recurrence equations that can be solved numerically in polynomial time on every graph. In this way, one can study very general assignments of  parameters and node functions $\{p_i, \mu_i, r_i, \eta\}$.

Here we present a derivation of these iterative equations that is based on a simple detailed balance argument for the single-node dynamics in the factorized approximation. A more precise derivation is presented in \ref{app2}, where the same equations are obtained by means of a series of approximation starting from very general exact results. 

In the single-node description, the transition rates for the queue length of node $i$ are 
\begin{eqnarray}\label{eq_tran}
w(n_{i} \to n_{i}+1) & = & p_i + (1-\mu_i) (1- \eta(n_i)) \sum_{j \in v(i)} \frac{r_j (1-\delta_{n_{j},0})}{k_{j}} \\
w(n_{i} \to n_{i}-1) & = &  \frac{r_i \theta(n_i)}{k_i} \sum_{j \in v(i)} \left[ 1- \eta(n_j) \right] 
\end{eqnarray} 
where $v(i)$ is the neighborhood of $i$.  A reasonable choice for the rejection probability is $\eta(n_i) = \bar\eta \theta(n_{i}- n^{*})$, where node $i$ is congested as soon as $n_i > n^*$. 
Imposing the detailed balance, the distribution $\mathcal{P}_{i}(n_{i})$ turns out to be a combination of exponentials, depending on $n_{i} > n^{*}$ or not (see also \ref{app2}). 

We expect three different behaviors: {\em free} nodes with exponentially decreasing distribution, {\em congested} nodes with exponentially increasing distribution (not normalizable), and unstable nodes, whose distribution is peaked around $n_i^*$, that we call {\em fickle} nodes.
Assuming the validity of the double-exponential single-node distributions, it is easy to show that the stationary state can be described in
terms of only two local quantities:  $q_{i} = Prob \left\{ n_{i} = 0 \right \} \in [0,1]$, the probability of empty queue at node $i$, and 
 $\chi_{i} = Prob \left\{ x_{i} = 1 \right\} \in [0,1]$,  the probability that node $i$ has more than $n^*$ particles ($x_i = \theta(n_i-n^*)$).

A congested node has always $n_{i} > n^{*}$, thus $\chi_{i} = 1$ and $q_{i} = 0$. 
Using Eqs. \ref{eq_tran}, the average rate of increase of $n_{i}$ in the stationary state is given by 
\begin{equation}\label{eq_ndot}
\dot{n}_{i} = p_i + (1-\mu_i)(1-\bar\eta \chi_{i}) \sum_{j \in v(i)} \frac{r_j(1-q_{j})}{k_{j}} - \frac{(1-q_{i}) r_i}{k_i} \sum_{j \in v(i)} [ 1- \bar\eta \chi_{j}] .
\end{equation}
It is straightforward to verify that $\dot{n}_{i} >0$ for congested nodes.\\
 Free nodes always have $n_{i} < n^{*}$, thus $\chi_{i} =0$. In addition $\dot{n}_{i} = 0$ otherwise they will eventually become congested. Imposing this condition on Eq. \ref{eq_ndot}, we get 
\begin{equation}\label{eq_q}
q_{i} = Q_{i}(\vec{\chi}, \vec{q}) \equiv 1- \frac{p_i+ (1-\mu_i) \sum_{j \in v(i)}\frac{r_j(1-q_{j})}{k_{j}}}{r_i-\frac{r_i}{k_{i}} \sum_{j \in v(i)} \bar\eta\chi_{j}} .
\end{equation} 
The case of fickle nodes is more tricky because their probability to be empty is expected to be exponentially small with $n^*$, i.e. $q_{i} \propto e^{-n_i^*}$, and vanishes only for $n^* \to \infty$. For the sake of simplicity we consider such a limit, 
that is approximately correct when $n^* \gg 1$. Hence, imposing $\dot{n}_{i}=0$ and neglecting $q_i \ll1$, we find an expression for  $\chi_{i}$,
\begin{equation}\label{eq_chi}
\chi_{i} = C_{i}(\vec{\chi},\vec{q}) \equiv \frac{1}{\bar\eta} \left[ 1 + \frac{ p_i - \frac{r_i }{k_{i}} \sum_{j \in v(i)} [1- \bar\eta \chi_{j}]}{(1-\mu_i) \sum_{j \in v(i)} \frac{r_j(1-q_{j})}{k_{j}}} \right].  
\end{equation}
It is easy to check that for a congested site $C_{i}(\vec{\chi},\vec{q})>1$ and $Q_{i}(\vec{\chi},\vec{q}) <0$, therefore the previous expressions can be written in the more compact way
\begin{equation}\label{iterative-eq}
\begin{array}{cc}
\chi_{i} & = \max \left\{ 0, \min \left[ 1, C_{i}(\vec{\chi},\vec{q})\right] \right\}, \\
q_{i} & = \max \left\{ 0, \min \left[ 1, Q_{i}(\vec{\chi},\vec{q})\right] \right\}.
\end{array}
\end{equation}  
These self-consistent equations can be solved iteratively on any specific graph. 

\begin{figure}
\begin{center}
\includegraphics*[width=0.45\textwidth]{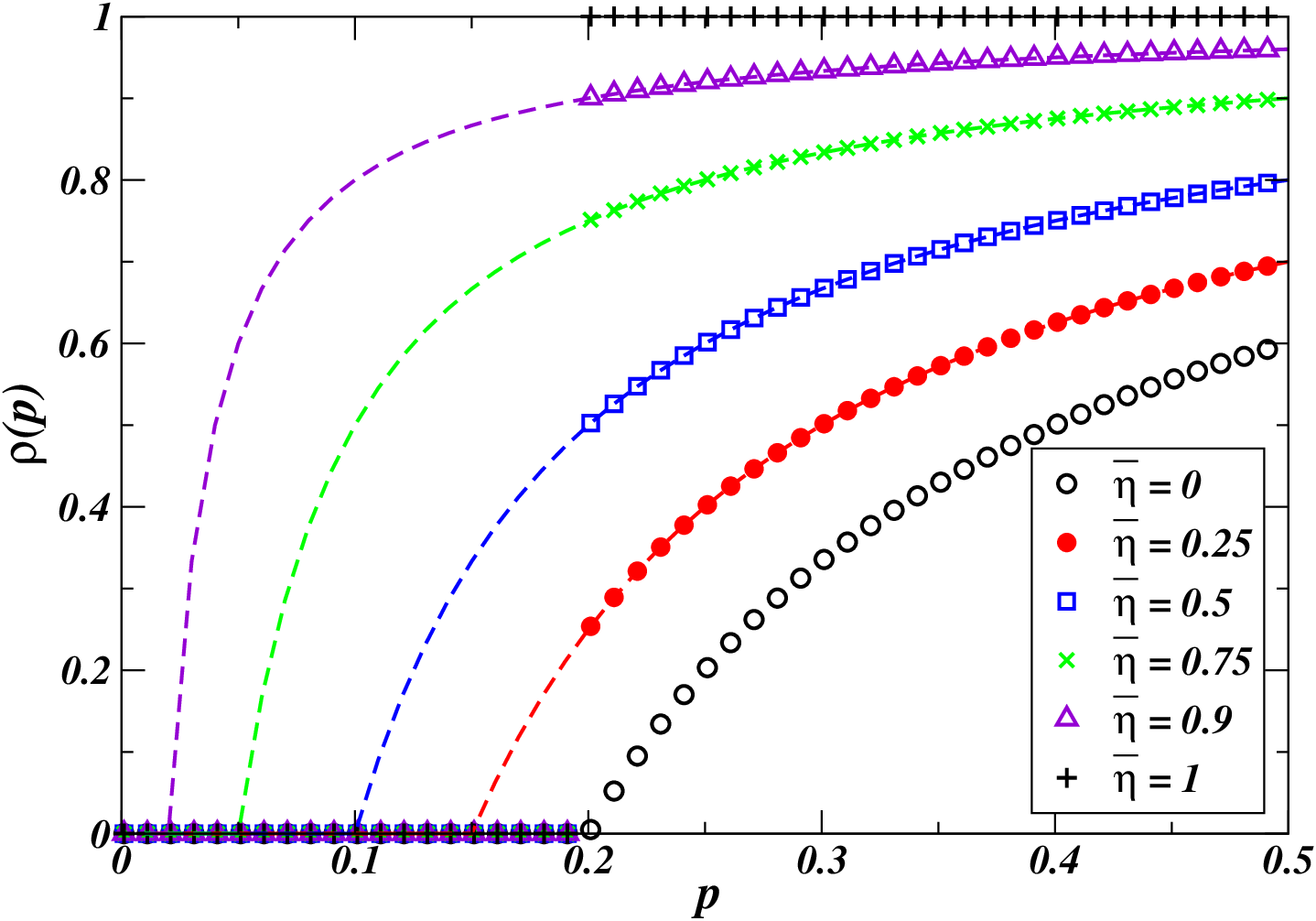}
\includegraphics*[width=0.45\textwidth]{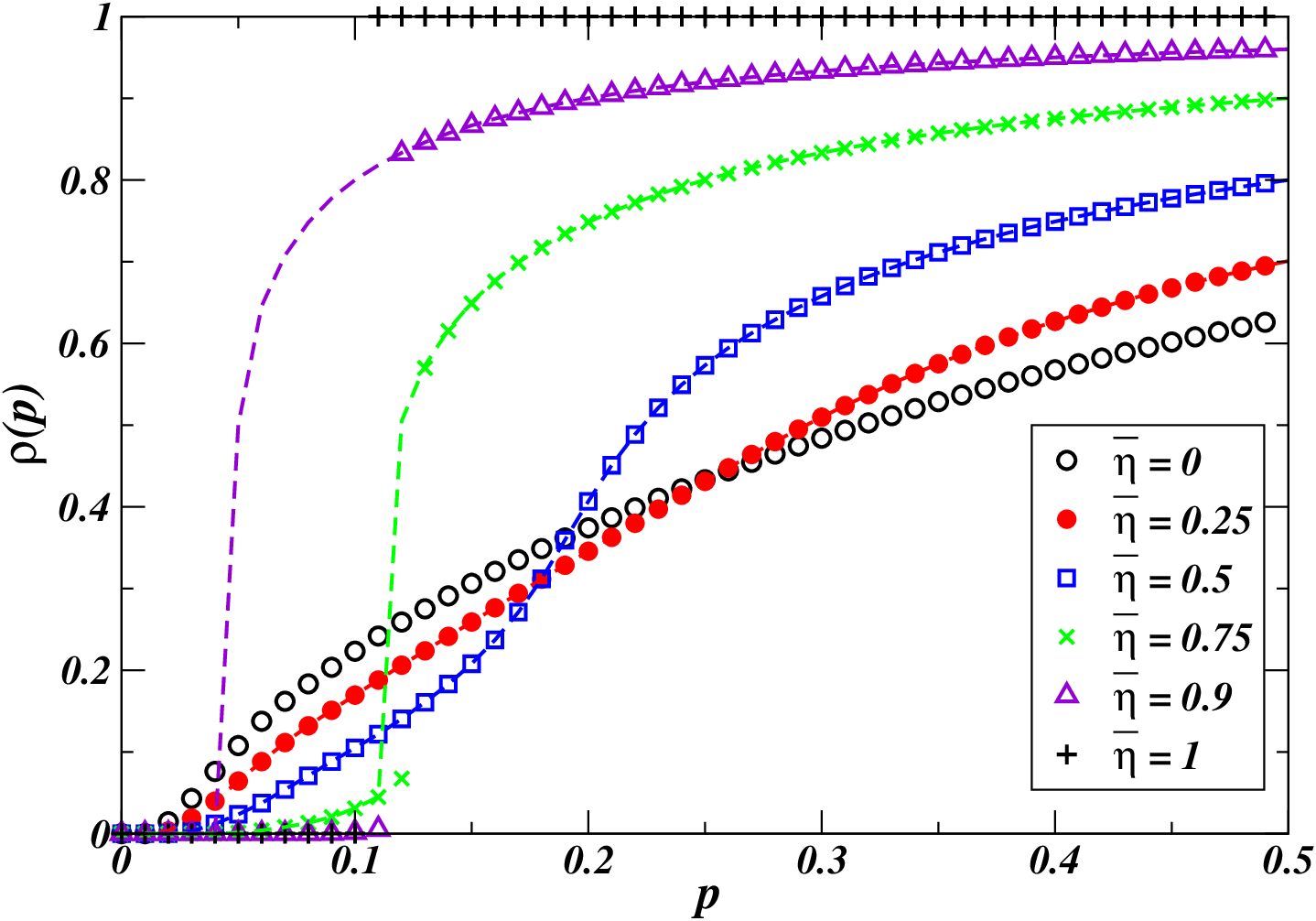}
\caption{
Behavior of the congestion parameter $\rho(p/\mu)$ for a random regular graph of size $N = 10^3$ and degree $K=4$ (left) and a scale-free network of size $N=10^3$ and exponent $\gamma = 2.5$. The symbols represents the behavior obtained by solving numerically the iterarive equations \ref{iterative-eq}  for increasing values of $p$ and $\eta = 0$ (black open circles), $0.25$ (red full circles), $0.5$ (blue squares), $0.75$ (green crosses) $0.9$ (violet triangles), $1$ (black crosses). Solving the Eqs. \ref{iterative-eq} for decreasing values of $p$, we find instead the corresponding dashed curves.
}
\label{fig-iter}
\end{center}
\end{figure}

If a fixed point of Eqs. \ref{iterative-eq} exists, the global congestion level can be measured by the order parameter $\rho = \frac{1}{p N} \sum_{i} \langle \dot{n}_{i} \rangle$, where $\langle \dot{n}_{i} \rangle$ is the average growth rate of queue $i$ computed on the fixed point values.\\
As an example, Fig. \ref{fig-iter} shows the diagram $\rho(p)$ obtained solving Eqs.\ref{iterative-eq} (in the limit $n^* \to \infty$) on a random regular graph (left)  and a scale-free network (right) for an homogeneous choice of the parameters ($p_i = p$, $\mu_i = \mu = 0.2$, and $\bar\eta = 0, 0.25, 0.5, 0.75, 0.9, 1$). In the random regular graph (left), the case $\bar\eta = 0$, corresponding to a ZRP with creation and absorption of particles, presents a clear signature of a continuous phase transition from a free-flow regime to a congested phase. When $\bar\eta >0$, we expect hysteresis phenomena, thus we first consider the solutions obtained for increasing values of $p$. More precisely, we find a fixed point for a value of $p$, we slightly change $p$ and we iterate the Eqs. \ref{iterative-eq} starting from such solution until we find another fixed point.  Increasing $p$ from zero, the congested phase appears abruptly, with a discontinuous behavior, at the same critical rate $p_c = \mu$ (independently of the value of $\bar\eta$). On the contrary, decreasing $p$ from the congested region, the systems undergoes a continuum transition to the free-flow phase (dashed lines) whose position decreases increasing $\bar\eta$.\\
In the scale-free network (right) the critical value $p_{c} \ll \mu$ for $\bar\eta=0$, but it is shifted towards higher values as soon as $\bar\eta >0$, as already observed in Fig. \ref{rhop} and found in \cite{EGM05,DDBM09}. 
It is worthy noting that in scale-free networks the position of the transition depends on the value of $\bar\eta$ and the curve $\rho(p)$  changes convexity for increasing values of $\bar\eta$. Finally, for $\bar\eta = 0.75$ the transition occurs in two-steps, first a continuous transition then a discontinuous one at slightly larger values of $p$. As expected, decreasing $p$ from the congested phase we observe hysteresis (dashed lines). We have verified performing the same calculation on other networks that the double transition is not always present and depends on both the value of $\bar\eta$ and the tail of the degree distribution.  

Though our results on single graphs reproduce the phenomenology of congestion observed in previous numerical simulations \cite{EGM05}, they also pose many new questions about the nature of the phase transitions and the mechanisms behind them.  For this reason, we have developed a complementary mean-field analysis at the level of random graphs ensembles that is able to shed light on all these points. 

\subsection{Mean-field analysis at the ensemble level}
\label{sec-theory2}

We consider uncorrelated random graphs with
degree distribution $P(k)$, so that $n_{k}$ represents now the average
queue length of nodes in classes of degree $k$. For the sake of simplicity, we will examine the simple homogeneous case $p_i = p$,
$\mu_i=\mu$, $r_i = 1$ and $\eta(n)=\bar\eta\theta(n-n^*)$.
We define $q_{k}=P\{n_i=0|k_i=k\}$ as the probability that a node of degree $k$ has
empty queue, and $\chi_{k}=\bar\eta P\{n_i\ge n^*|k_i=k\}$ as the probability that a node of degree $k$ refuses particles
(it should be noticed that now $\chi$ has an $\bar\eta$ factor). 
The mean-field transition rates for nodes with degree $k$ are
\begin{eqnarray}
w_{k}(n \to n+1) &=& p + (1-\mu)(1-\bar{q})\frac{k}{z}(1-\bar\eta\theta(n-n^{*}))\nonumber \\
w_{k}(n \to n-1) &=& \theta(n)(1- \bar{\chi}),
\end{eqnarray}
where $z$ is the average degree,  $\bar{q} = \sum_{k} q_{k} P(k)$ and $\bar{\chi} = \sum_{k} \frac{k}{z} \chi_{k} P(k)$.
The average queue length $\langle n_{k}\rangle$ follows the rate equation
\begin{equation}\label{dotn}
\langle\dot{n}_{k} \rangle = p +(1-\mu) (1-\bar{q}) \frac{k}{z} (1 - \chi_{k})-(1-q_{k}) (1-\bar{\chi}) .
\end{equation}
Note that summing over $k$ and dividing by $p$ we obtain a measure of the order parameter $\rho (p)$. 

Since $\dot{n}_{k}$ depends linearly on $k$, high degree nodes are more likely to be congested, 
therefore, for every $p$, there exists a real valued threshold  
$k^{*}(p)$ such that all nodes with $k > k^{*}$ are congested whereas nodes with degree less than $k^*$ are not congested.
Congested nodes ($k > k^{*}$) have $q_{k} = 0$ and $\chi_{k} =\bar\eta$.
The probability distribution for the number of particles in the queue of
free nodes with degree  $k<k^{*}$ can be extracted by calculating
the generating function $G_{k}(s) = \sum_{n} \mathcal{P}_k(n_k=n)s^n$ from the
detailed balance condition $w_k(n_k+1 \to n_k) \mathcal{P}_k(n_k+1) = w(n_k \to n_k+1) \mathcal{P}_k(n_k)$, that we assume to hold in this approximation (see also \ref{app2}).
 The generating function takes the form
\begin{equation} \label{genfun}
G_{k}(s) = q_{k}\left\{ \frac{1-{(a_{k} s)}^{n^{*}}}{1-a_{k} s} +
  \frac{{(a_{k} s)}^{n^{*}}}{1-(a_{k} - b_{k})s}\right\}
\end{equation}
 corresponding to a double exponential, where $a_{k} = [p+(1-\mu) \frac{k}{z} (1-\bar{q})]/[1-\bar{\chi}]$ 
and $b_{k} = \bar\eta [(1-\mu)\frac{k}{z}(1-\bar{q})]/[1-\bar{\chi}]$. 
From the normalization condition $G_{k}(1)=1$ and the condition $\dot{n}_{k} = 0$, 
we get expressions for $q_k$, $\chi_k$,
\begin{eqnarray}
q_k & = & \left[  \frac{1-a_{k}^{n^{*}}}{1-a_{k}} +
  \frac{a_{k}^{n^{*}}}{1-a_{k} + b_{k}} \right]^{-1} \\
  \chi_k & = & 1 + \frac{p-(1-q_k)(1-\bar\chi)}{(1-\mu)(1-\bar{q})\frac{k}{z}}
\end{eqnarray}
and, finally, for $\bar{q}$, $\bar{\chi}$.

The value $k^{*}$ is  self-consistently determined imposing that nodes with $k=k^{*}$ 
are marginally stationary, i.e. $\dot{n}_{k^{*}}=0$ with $q_{k^{*}} = 0$, $\chi_{k^{*}} =\bar\eta$, that translates into the equation
\begin{equation} \label{kstar}
k^{*} = \frac{1-p-\bar{\chi}}{(1-\mu)(1-\bar\eta)(1-\bar{q})}z . 
\end{equation}
The set of closed equations for $\bar{q},\bar\chi$ can be solved numerically for any degree distribution 
$P(k)$ and $\rho(p)$ can be accordingly computed.

\subsubsection{Homogeneous Networks --}
The equations for $\bar{q}$ and $\bar\chi$ simplifies to a single equation when all nodes have the same properties, and in particular the same degree  ($k_i=K, ~\forall i$). On these networks, the mean-field behavior can be trivially studied for any value of $n^*$, but we consider as an illustrative example the limit $n^{*}\to \infty$. 
Only two solutions of the equation relating $\bar{q}$ and $\bar\chi$ are possible: the free-flow solution ($\rho=0$) with $\bar q=1-p/\mu$ 
and $\bar\chi=0$ that exists for $p\le \mu$, and congested-phase solution, where all nodes have 
$n_i\to\infty$, i.e. $\bar\chi=\bar \eta$ and $\bar q=0$. The latter solution has $\rho=\dot n/p=1-(1-\bar\eta)\mu/p$ 
and exists for $p\ge (1-\bar\eta)\mu$. 
There is a simple argument explaining the law $1- (1-\bar\eta)\mu/p$, that corresponds to the situation in which the whole system is congested. 
In the infinitesimal interval of time $dt$, $p N$ particles are created, and $N$ particles try to hop. The nodes are congested, so a fixed fraction $\mu(1-\bar\eta)$ of them is absorbed, and the expression of $\rho(p)$ follows from its definition (Eq.\ref{op}). 

\begin{figure}[h]
\begin{center}
\includegraphics*[width=0.5\textwidth]{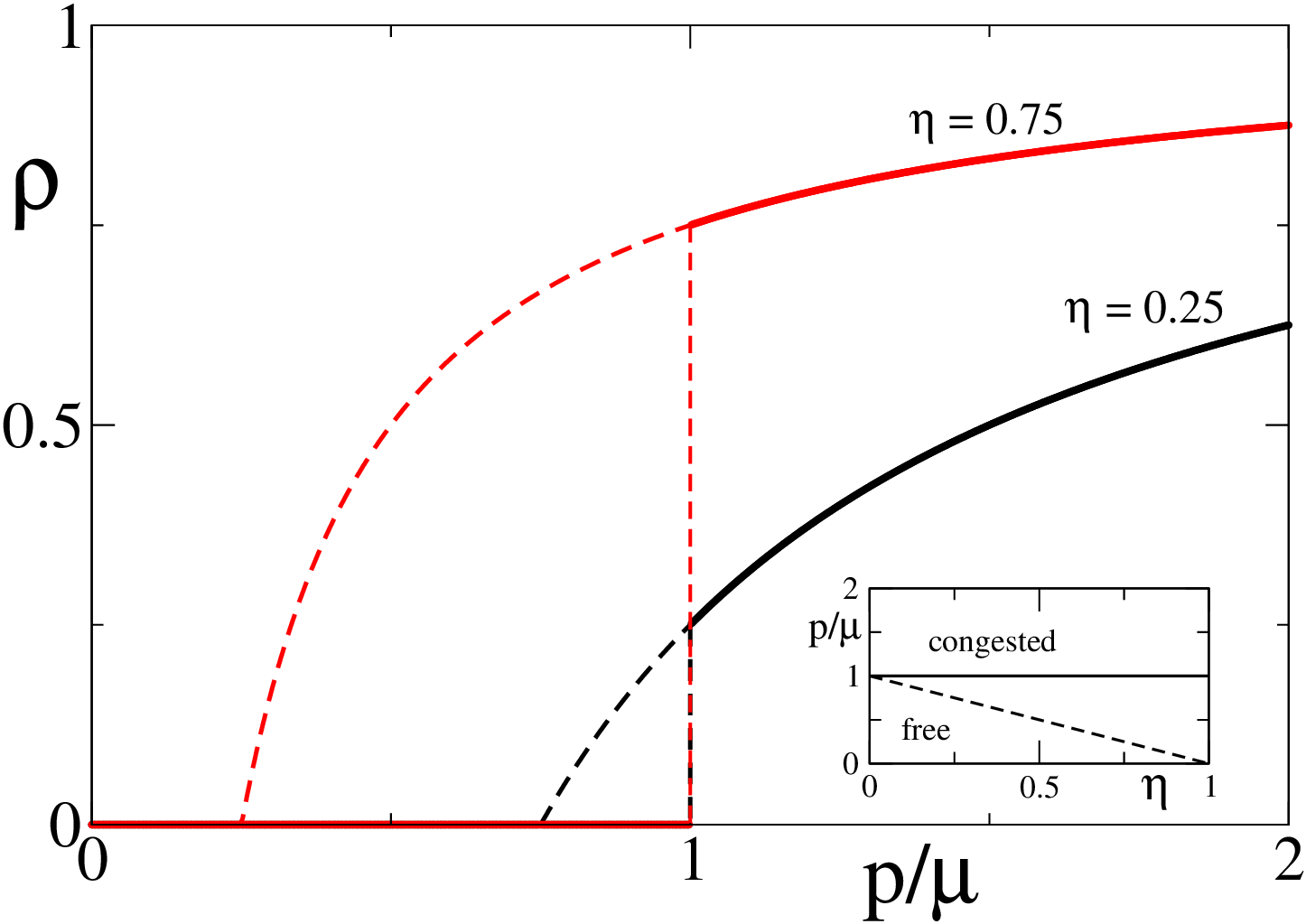}
\caption{Behavior of the congestion parameter
$\rho(p/\mu)$ for a random regular network obtained theoretically 
for $\eta = 0.25$,  $0.75$. Inset: phase diagram for the same graph.
}
\label{figreg}
\end{center}
\end{figure}

The behavior of the congestion parameter with both the continuous and discontinuous transitions to the congested state  is plotted in Fig. \ref{figreg} for $\bar\eta=0.25, 0.75$.
The corresponding phase diagram, reported in the inset of Fig. \ref{figreg}, shows that in the interval $p\in [(1-\bar\eta)\mu,\mu]$ both a congested- and a free-phase coexist. We find  an hysteresis cycle, with  the system that turns from a free phase into
 a congested one discontinuously as $p$ increases and crosses $p=\mu$, and it reverts back to the free phase 
only at $p=(1-\bar\eta)\mu$ as $p$ decreases. 

\subsubsection{Heterogeneous Networks --}
\begin{figure}[h]
\begin{center}
\includegraphics*[width=0.5\textwidth]{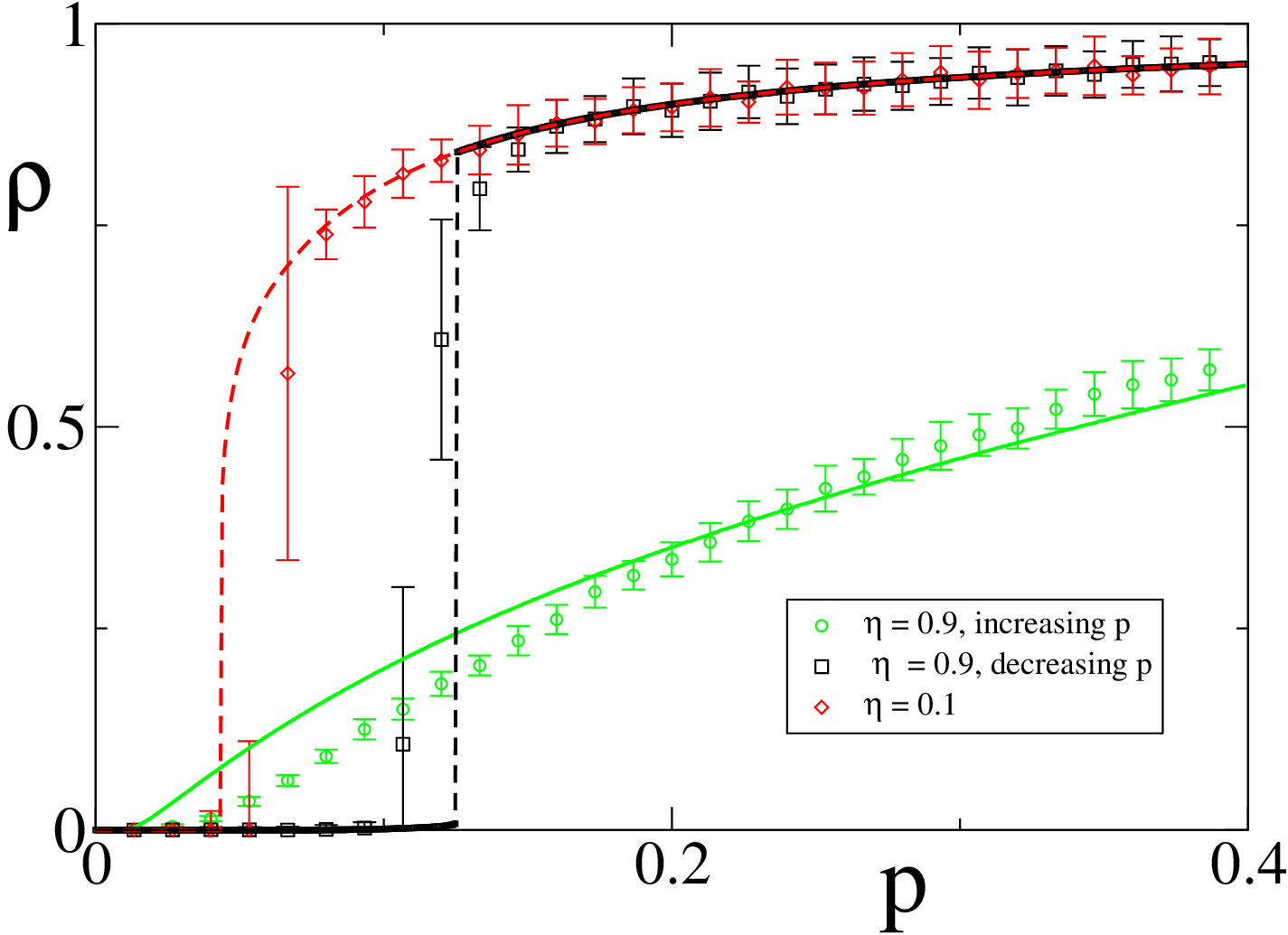}
\caption{
$\rho(p)$ for an uncorrelated scale-free graph
($P(k)\propto k^{-3}$, $k_{min}=2$, $k_{max} = 110$, $N = 3000$),
$\mu=0.2$, $n^{*}=10$ and $\bar\eta=0.1$
and $\bar\eta=0.9$, from both simulations (points) and theoretical predictions (lines). Hysteresis is observed increasing (black curve and points) and then decreasing (red curve and points) $p$ across the transition.
}
\label{fig1}
\end{center}
\end{figure}

In the case of heterogeneous networks the equations for $\bar{q}$ and $\bar\chi$ have to be solved numerically. 
For instance, in Fig. \ref{fig1} we compare the theoretical 
prediction (full line) for $\rho(p)$ in a scale-free network  with results of  simulations (points). 
The agreement is good, the theoretical prediction at the ensemble level  confirming the scenario already observed in the simulations of Section \ref{sec-model2} and in the single-graph analysis of Section \ref{sec-theory1}. 
The curves are obtained for $\mu = 0.2$ and $n^{*} = 10$, but the behavior does not 
qualitatively change for different values of these parameters. The dependence on $\bar\eta$ brings instead qualitative changes.
Increasing $\bar\eta$ from $0.1$ to $0.9$, the transition becomes discontinuous and $p_{c}$ increases. 

\begin{figure}[h]
\begin{center}
\includegraphics*[width=0.35\textwidth]{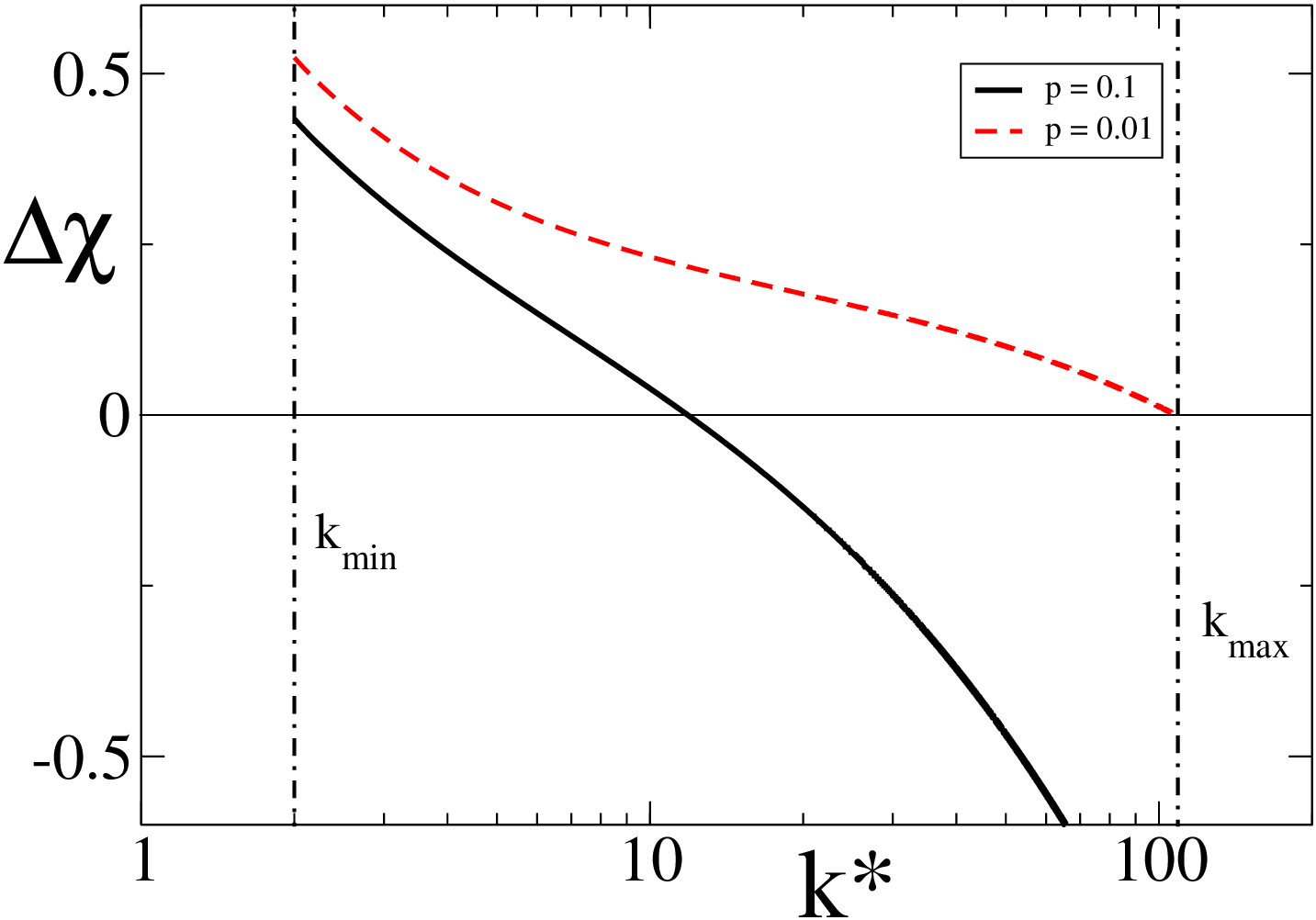}
\includegraphics*[width=0.35\textwidth]{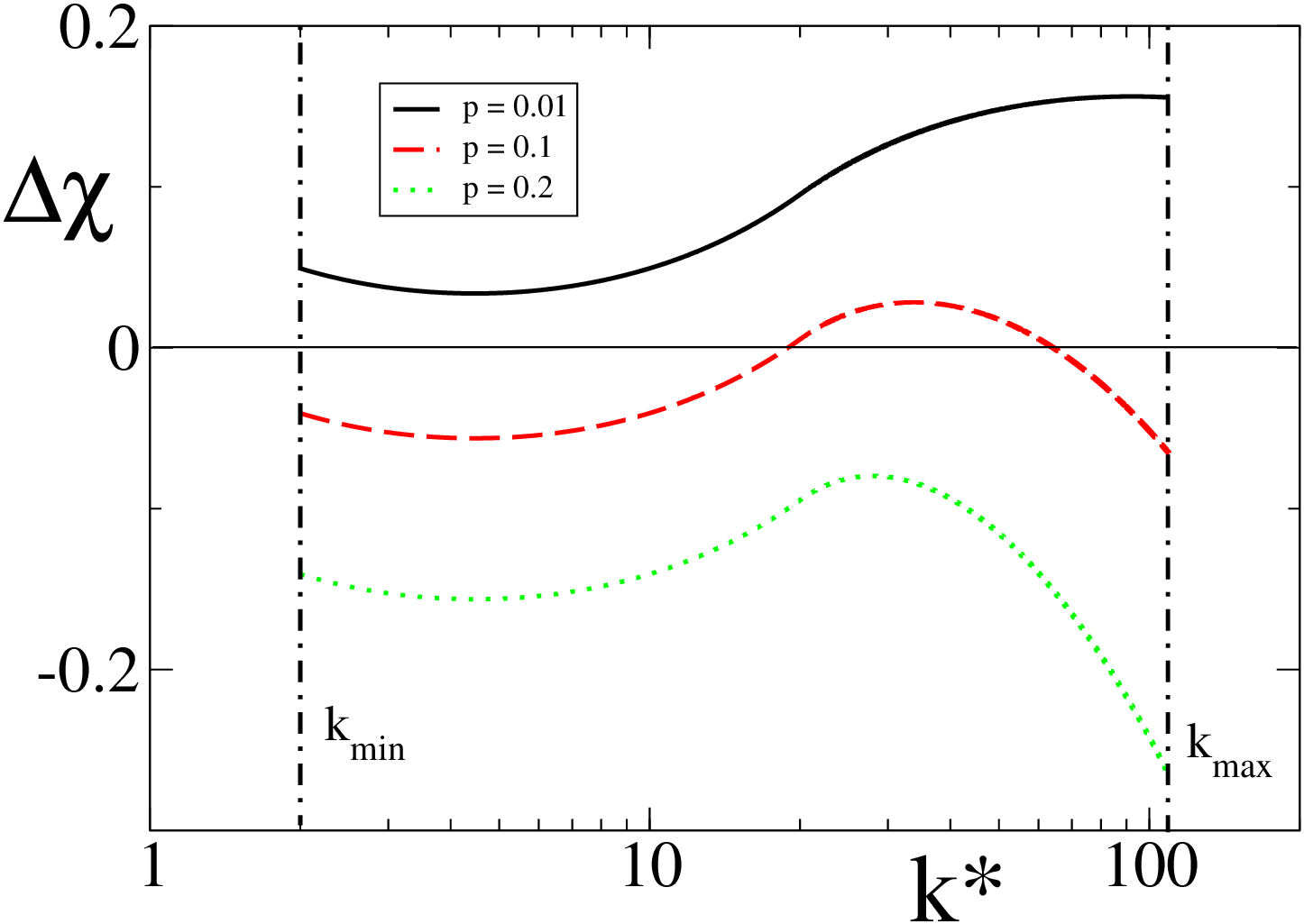}
\caption{
The zeros of $\Delta \chi (p)$ vs. $k^{*}$ define the threshold degree for the onset of congestion in a network.
The picture refers to a scale-free random network with $\gamma = 3.0$, $k_{min}=2$ and $N=3000$
($k_{max}= 110$), and different values for $\bar\eta = 0.1$ (left) and $0.9$ (right) and $p$. The
solution $k^*_1(p)$ in the right panel falls outside the plot.
} \label{fig2}
\end{center}
\end{figure}

\begin{figure}[hb]
\begin{center}
\includegraphics[width=0.4\textwidth]{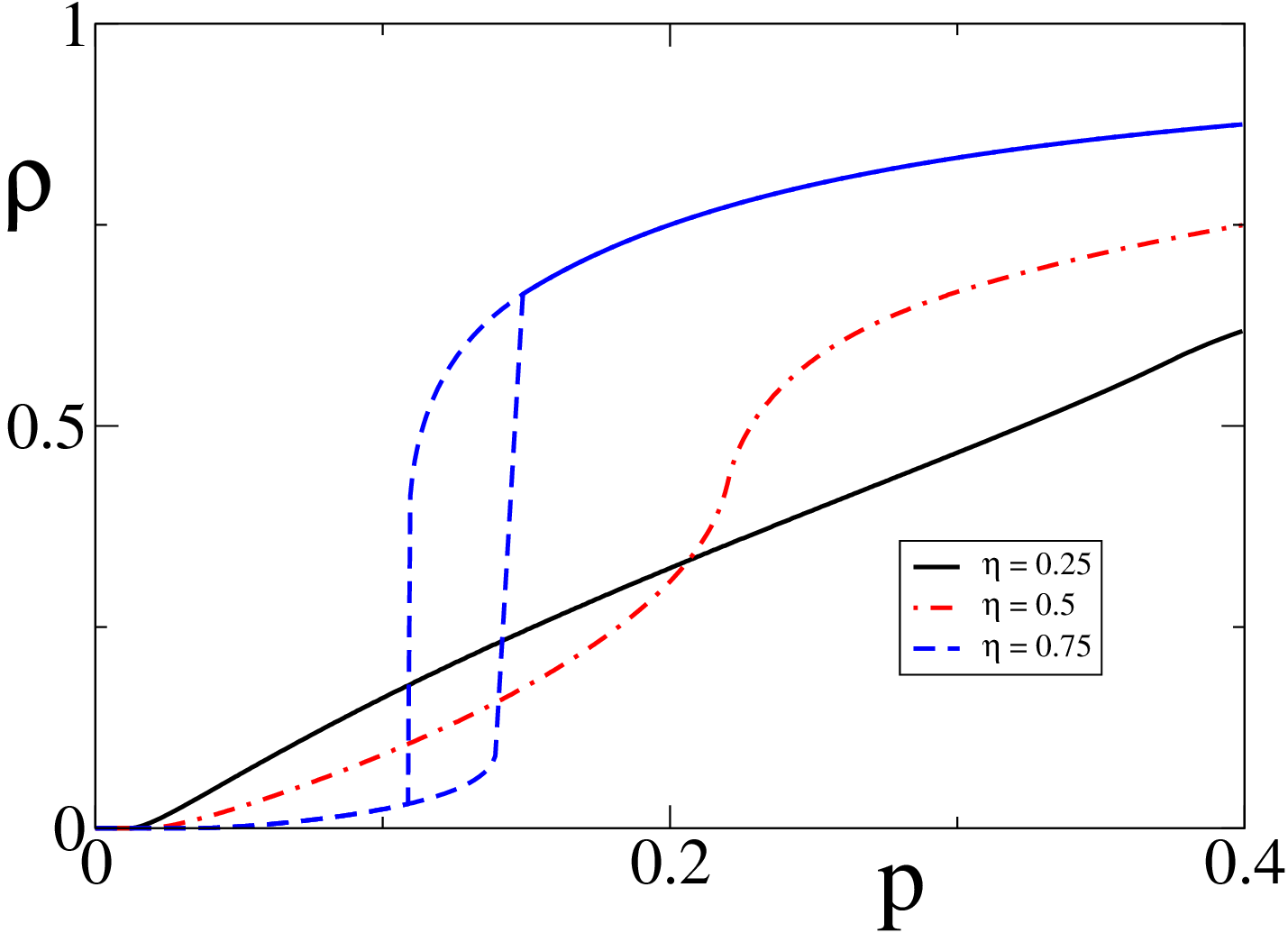}
\caption{Increasing $\bar\eta$, the congestion parameter $\rho(p)$ develops a discontinuous transition. Here we report the case of the graph of Fig.\ref{fig2}. For $\eta=0.75$, we have first a continuous, then a discontinuous transition.  
} \label{fig_OP}
\end{center}
\end{figure}

The main difference with respect to homogeneous networks is that on heterogeneous networks, not all nodes become congested at the same time. The rate $p$ at which a node becomes congested depends on its degree, the hubs being first. The process governing the onset of congestion and the effects of the rejection term can be understood in the limit $n^{*} \to \infty$, that  simplifies considerably the calculations without modifying the overall qualitative behavior for sufficiently large $n^*$. We refer the reader to \ref{app3} for a discussion of the main differences emerging in the case of low values of $n^*$.  \\ 
We have to solve in the limit $n^* \to \infty$ the self-consistent equations for $\bar\chi$ and $\bar{q}$. In this limit, uncongested nodes have $a_{k} <1$, hence  $\chi_{k} \to 0$ and $q_{k} = 1- a_{k}$. All nodes with degree $k<k_F$, where  $k_{F} = \max (k^{*} (1-\bar\eta), k_{min})$, are free from congestion.  Congested nodes have $q_{k} \to 0$ and $\chi_{k} = \bar\eta$ (for $k \geq k^*$). The fickle nodes are those with $k_F \leq k < k^{*} $ and they have $\chi_{k} = 1-\frac{k_F}{k}$. 
Using this classification, we get a first expression for $\bar\chi$, i.e.
\begin{equation}
\bar{\chi}_{1}  = \sum_{k = k_{F}}^{k^{*}} \left[ 1 - \frac{k_F}{k}\right]\frac{k}{z} P(k) 
+ \bar\eta \sum_{k=k^{*}}^{k_{max}} \frac{k}{z} P(k) .
\end{equation}
Eq. (\ref{kstar}) provides a further relation between $\bar{q}$, $\bar{\chi}$ and $k^*$. We eliminate $\bar{q}$  using its definition which leaves us with another expression for $\bar{\chi}$,
\begin{equation}
\bar{\chi}_{2} = 1 - \frac{1}{2A}\left\{ 1+ A p -B + {\left[ (1+ A p -B)^2 + 4 A B p \right]}^{1/2}\right\} 
\end{equation}
where $A = z/[k^{*}(1-\bar\eta)(1-\mu)]$ and $B = \sum_{k = k_{min}}^{k_{F}} \left[ 1- \frac{k}{k_F}\right] P(k)$. To determine $\bar{\chi}$ we have to solve the implicit equation $\bar\chi_1 = \bar\chi_2$.\\
In Fig. \ref{fig2} we plot the difference $\Delta \chi = \bar\chi_{1} - \bar\chi_{2}$ vs. $k^{*}$, for $\bar\eta=0.1$ (left) and $0.9$ (right) and different values of $p$ on a scale-free graph.
The zeros of $\Delta\chi(k^*)$ correspond to the only possible values assumed by $k^{*}$. For small rejection probability ($\bar\eta = 0.1$ in Fig. \ref{fig2}), there is only one solution $k^{*}(p)$, which decreases from  $+\infty$ when increasing $p$ from $0$. The value $p_{c}$ at which $k^{*}(p_{c}) = k_{max}$ is the critical creation rate at which largest degree nodes become congested. At larger $p$, $k^{*}(p)$  decreases monotonously until eventually all nodes are congested when $k^{*}(p) = k_{min}$.  Hence for low values of $\bar\eta$, the transition from free-flow to the congested phase occurs continuously at the value of $p$ for which $k^{*}(p) = k_{max}$. \\
At large $\bar\eta$ ($\bar\eta = 0.9$ in Fig. \ref{fig2}), the scenario is more complex. Depending on $p$, the equation can have up to three solutions, $k^{*}_{1} (p)\leq k^{*}_{2} (p)\leq k^{*}_{3}(p)$. It is easy to check that only $k^{*}_{1}$ and $k^{*}_{3}$ can be stable solutions. For $p\ll 1$ there is only one solution at $k^{*}_{3} (p)\gg k_{max}$, corresponding to the free phase. This is thus the stable solution for $p$ increasing from zero. As $p$ increases, another solution $k^*_1(p) < k^*_3(p)$ can appear, and $k^*_3(p)$ moves towards lower degree values. Three situations may occur:
\begin{itemize}
\item[{\em i.}] The solution $k^{*}_{3}(p)$ disappears before reaching $k_{max}$. Then $k^*_1(p)$ becomes the stable solution, and the congested phase appears abruptly.  However, given the shape of the function $\Delta \bar\chi$ (see Fig. \ref{fig2}), when this happens $k^*_1(p) \to 0$ and in particular we expect $k^*_1(p) < k_{min}$, so that  above the transition the whole network is congested and follows the law  $\rho(p) = 1-(1-\bar\eta)\frac{\mu}{p}$.
\item[{\em ii.}] The solution $k^{*}_{3}(p)$ crosses $k_{max}$ and exists until it reaches $k_{min}$. Then the congested phase emerges continuously and the network is only partially congested (i.e. only the nodes with $k \ge k^*_3(p)$). The order parameter grows until it reaches the curve of complete congestion $\rho(p) = 1-(1-\bar\eta)\frac{\mu}{p}$ ($k^*_3(p) < k_{min}$).
\item[{\em iii.}] The solution $k^*_3(p)$ crosses $k_{max}$ but disappears before reaching  $k_{min}$, and $k^*_1$ becomes the stable solution. In this case the congested phase appears continuously (only high-degree nodes are congested), but at some point another transition occurs that brings the system abruptly into the completely congested state.
\end{itemize}
The possible situations are summarized in Fig.\ref{fig_OP}, where we have sketched the corresponding behaviors of the order parameter $\rho(p)$. 

The scenario at points {\em i.-ii.} is exactly that of Fig.\ref{fig1}, while a signature of the double-transition can be observed in Fig.\ref{fig-iter} (right). In general, the exact phenomenology observed in numerics and simulations depends strongly on the tail of the degree distribution, i.e. on the graph ensemble considered. 

Note that in case of discontinuous transitions, the presence of an hysteresis phenomenon is associated to the stability of the two solutions $k^*_{1}(p)$ and $k^*_{3}(p)$. For instance, in case {\em ii} or {\em iii}, we start from the free-phase at low $p$, the system selects the solution $k^{*}_{3}(p)$ and follows it upon increasing $p$ until the solution $k_3^*(p)$ disappears. On the contrary, starting from the congested phase (large $p$) the system selects the solution $k_1^*(p)$ and remains congested until this solution disappears (see inset of Fig. \ref{fig1}). 

\begin{figure}[h]
\begin{center}
\includegraphics[width=0.45\textwidth]{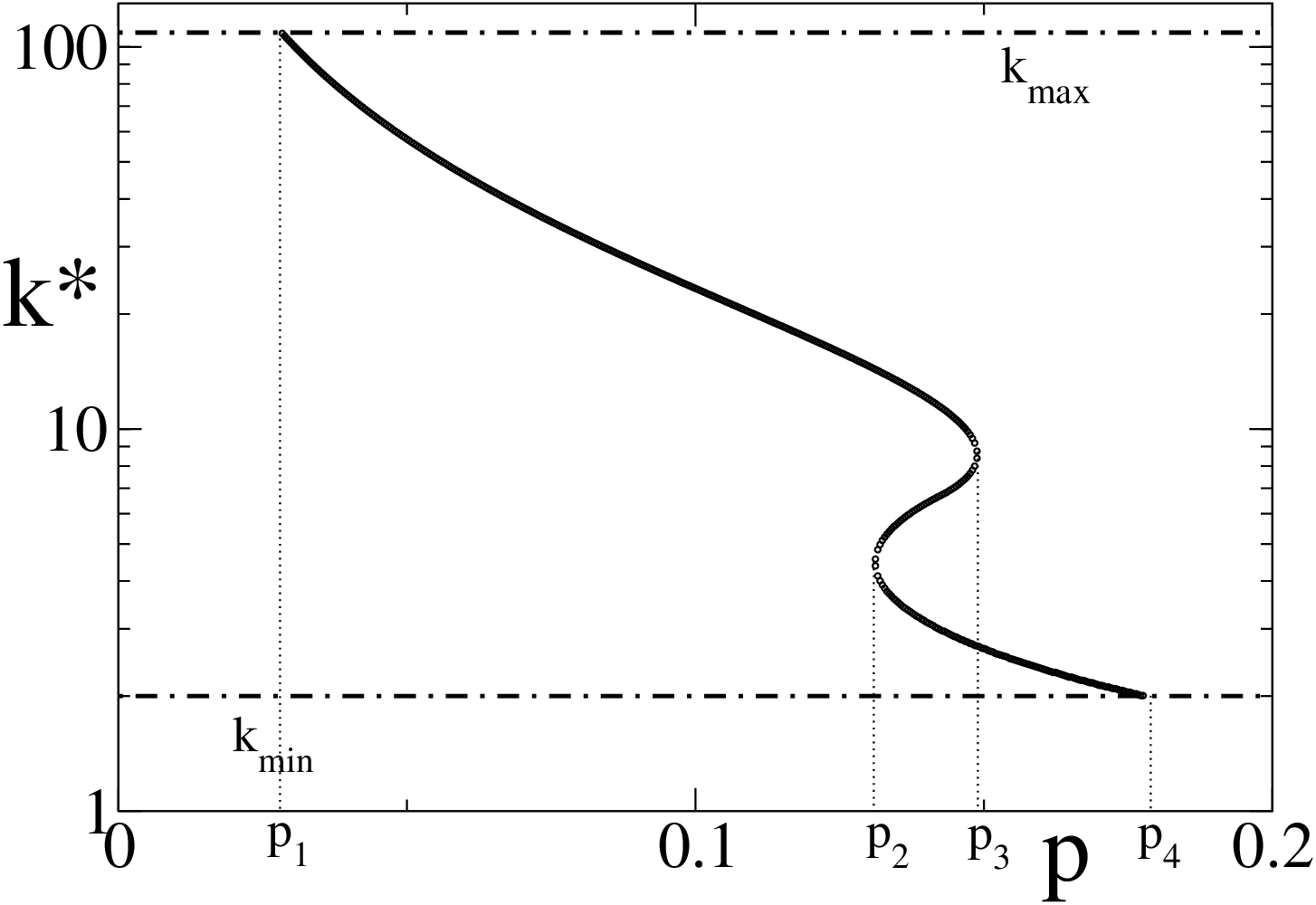}
\caption{The solution $k^*(p)$ for the scale-free graph of Fig.\ref{fig1}, with $\bar\eta=0.7$. 
At $p_1$, $k^*=k_{max}$, and the system becomes partially congested in a continuous way. 
Between $p_2$ and $p_3$ there are three solutions, two of them are stable.
Increasing $p$, the system jumps suddenly to a more congested state at $p_3$, whereas decreasing $p$, the system jumps to a less congested state at $p_2$. Above $p_4$ the system is completely congested. 
} \label{keta}
\end{center}
\end{figure}

In Fig. \ref{keta} we can see the solution $k^*(p)$ for the same graph of Fig.\ref{fig1}, with $\bar\eta=0.7$:
at $p_1$, when $k^*=k_{max}$, the system becomes congested in a continuous way, at $p_3$ there is a discontinuous jump
to higher values of congestion, while above $p_4$ the network is fully congested and finally, coming back to $p_2$ 
there is a jump to a less congested state. Between $p_2$ and $p_3$ there is coexistence of high
and low congested states with hysteresis.\\

In summary, the system can show a sort of hybrid transition: a continuous transition to a partially congested state followed by a discontinuous one to a (almost) completely congested one (see Fig.\ref{fig_OP}).

\subsubsection{The General Phase-Diagram}
On heterogeneous random graphs, the behavior of the system in the plane $(\bar\eta, p)$ depends in a complex way on its 
topological properties, such as the degree cut-off and the shape of the degree distribution. For this reason the precise location of the
critical lines, separating different phases, can be determined only numerically using the methods exposed in the previous section.
In the following, we give a qualitative description of the general structure of the phase diagram in the limit $n^* \to \infty$, then we substantiate the analysis reporting 
an example of phase diagram obtained numerically for the same networks ensemble of Fig. \ref{fig1}. 
 
%
%Fig. \ref{fig3} reports the phase diagram for the same uncorrelated 
%scale-free random networks considered in Fig. \ref{fig1} in the limit $n^{\star} \to \infty$. 
%The dashed line represents the continuous phase transition, separating free-flow regime from congestion. 
%At the point $C$, the critical line splits in two branches that define a coexistence region.
%The upper full line represents the discontinuous transition from the free-phase to the jammed state, 
%whereas the lower indicates the opposite continuous transition from congestion back to free-flow. 
%The dotted line decreasing from the maximum of the critical line, is an unphysical branch of the analytic solution. 
%Indeed, a free stationary state cannot become congested if we increase $\bar\eta$, because it affects only already congested nodes.

A first important region of the space of parameters is the one in which a completely free solution exists, i.e. $k_{max} \le k_F$. This solution is characterized by $\bar{q} = 1-p/\mu$, $\bar\chi = 0$ and $\rho=0$. From the expression for $\dot{n}_k=0$ computed in $k_{max}$ we find that this happens as long as  $p\le p_{c_0}$ with 
\begin{equation} p_{c_0} = \frac{\mu}{\mu + (1-\mu)\frac{k_{max}}{z}} . \end{equation}
Note that this region does not depend on the rejection probability $\bar\eta$, because rejection affects only congested nodes.

\begin{figure}[h]
\begin{center}
\includegraphics*[width=0.45\textwidth]{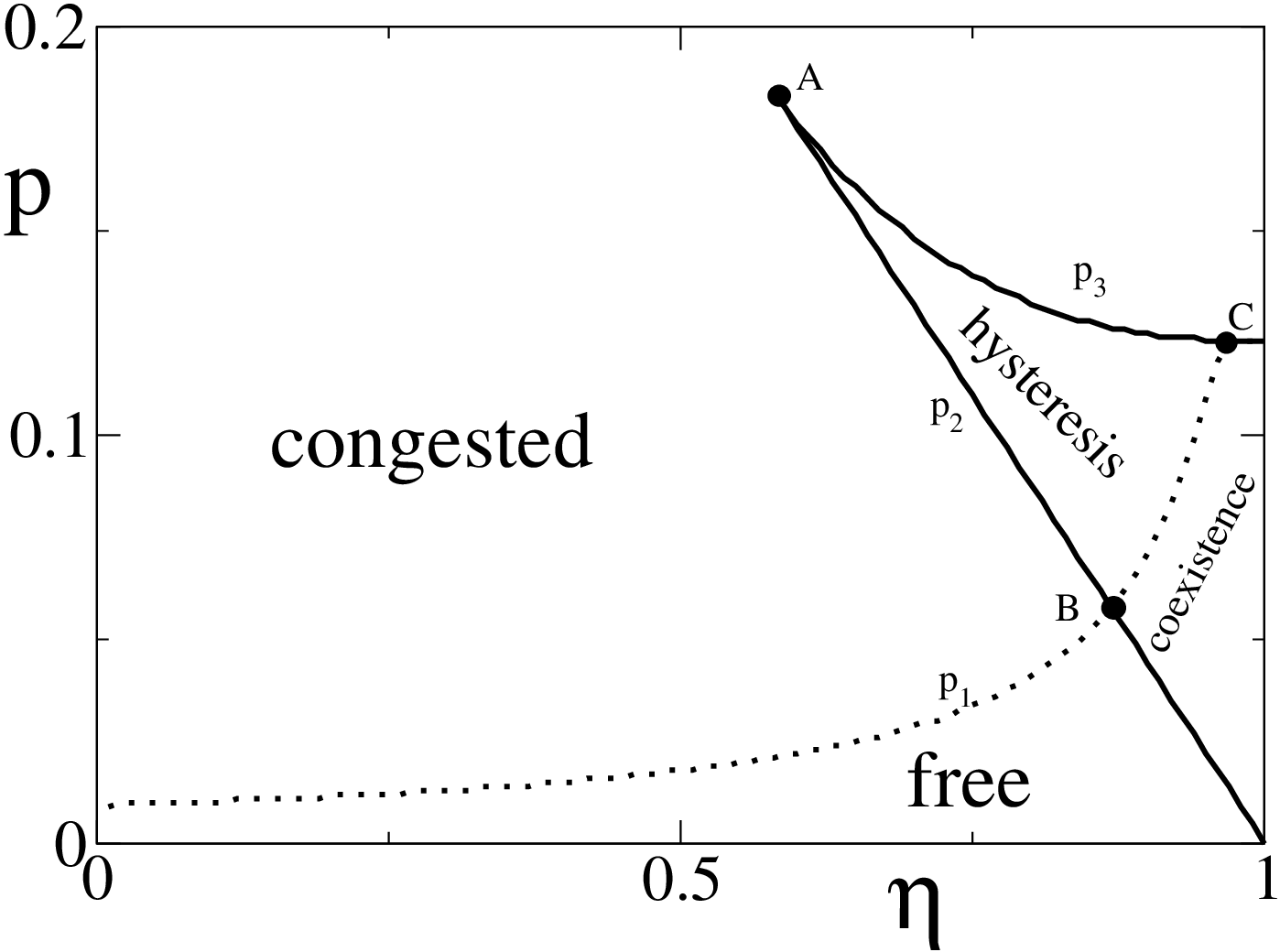}
\caption{
$(\bar\eta,p)$ phase diagram for the uncorrelated scale-free graph
of Fig. \ref{fig1}.
} \label{fig3}
\end{center}
\end{figure}

The transition takes place when the maximum degree nodes first become congested, i.e. $k^{*}=k_{max}$. Since $\dot{n}_{k^*} =0$, $q_{k_{max}}=0$ and $\chi_{k_{max}} = \bar\eta$, we get from Eq. \ref{dotn} a first expression for $p_{c} =  1-\bar\chi - \frac{k_{max}}{z}(1-\mu)(1-\bar\eta)(1-\bar{q})$. Now computing $\rho$ averaging Eq. \ref{dotn} and imposing  $\rho = 0$, we find a second expression for $p_{c} = \mu (1-\bar{q})(1-\bar\chi)$. Eliminating $\bar{q}$ from these two equations, we find the critical line
\begin{equation}
p_c(\bar\eta)=\frac{(1-\bar\chi)^2}{1-\bar\chi-\frac{k_{max}}{z}(1-\bar\eta)\frac{1-\mu}{\mu}}
\end{equation}
where $\bar\chi = \sum_{k\ge k_F} \frac{k P(k)}{z}\left(1-\frac{k_{max}(1-\bar\eta)}{k}\right)$.
Below this line (dotted line in Fig. \ref{fig3}) the system is not congested ($\rho = 0$), even if in the region $p_{c_0} \le p \le p_{c}(\bar\eta)$ higher-degree nodes are unstable ($k_F \le k_{max} \le k^*$). \\
It is possible to show that $p_c(\bar\eta)$ attains its maximum in $\bar\eta_c = 1-\frac{k_{min}}{k_{max}}$ where $p_{cmax} = \mu \frac{k_{min}}{z}$, where
$k_F=k_{min}$ and so above this point the curve is constant $p_c(\bar\eta>\bar\eta_c)=p_c(\bar\eta_c)$.

The transition line $p_c(\bar\eta)$ corresponds to the point $p_1$ in Fig. \ref{keta}, calculated for all values of $\bar\eta$. 
We can calculate the two curves $p_2(\bar\eta)$, $p_3(\bar\eta)$ as well, in order to get the points at which there are discontinuous jumps in the congestion parameter $\rho(p)$.

Looking at Fig. \ref{fig3} we can distinguish three points A, B, C dividing the phase diagram into different regions:
\begin{itemize}
\item[{\em i.}] Below $\bar\eta_A$ we have a continuous transition to a congested state increasing $p$ above $p_1$.
\item[{\em ii.}] Between $\bar\eta_A$ and $\bar\eta_B$ the transition is continuous at $p_1$. Then, increasing $p$ above $p_3$, 
there is a discontinuous jump to a more congested state. Coming back to lower values of $p$, there is a discontinuos jump
to a less but still congested state at $p_2$, and the system eventually becomes free below $p_1$ in a continuous way.
\item[{\em iii.}] Increasing $p$ in the region between $\bar\eta_B$ and $\bar\eta_C$, there is a smooth transition
from free-flow to a congested state at $p_1$, and a sudden jump to a more congested phase at $p_3$; but, this time, by
decreasing $p$ from the congested state, the transition to the free phase is discontinuous and located in $p_2$.
\item[{\em iv.}] For $\bar\eta > \bar\eta_C$ the transition is a purely discontinuous one with transition points $p_2$ and $p_3$.
\end{itemize}

Increasing $p$ above the transition, at some point $p_{c_1}(\bar\eta)$ the system becomes completely congested. For $p \ge p_{c_1}(\bar\eta)$, the order parameter follows the curve $\rho = 1-\mu(1-\eta)/p$. This happens for 
 $p \ge p_{c_1}(\bar\eta) = (1-\bar\eta)(1-(1-\mu) k_{min}/z)$, where $k^* \le k_{min}$, $q=0$, $\chi=\bar\eta$.

These calculations show that the phase diagram crucially depends on the tail of the degree distribution. In scale-free networks $k_{max}$ scales with the network's size $N$ as $N^{\frac{1}{\omega}}$ with $\omega = 2$ (structural cut-off) or $\omega = \gamma-1$ (natural cut-off). Accordingly the critical line depends on the system's size, $p_{c} \propto N^{-\frac{1}{\omega}}$. The only region that does not depend on $k_{max}$ is that after the maximum of the curve ($\bar\eta \ge \bar\eta_C$).

\subsection{Relation to condensation phenomena}
\label{sec-theory3}

We have discussed the stationary dynamics of the non-conserved model, but it is still not clear 
if the mechanism triggering the congestion phase transition is the same causing condensation in ZRP \cite{EH05}.  

The steady-state properties of the conserved model  ($p=0$, $\mu=0$) are exactly solvable once we have factorized the distribution
on the nodes. Imposing the detailed balance we get (the calculation with $r_i \neq 1$ is in \ref{app1}),
\begin{equation}
\mathcal{P}_i(n) = \left\{ \begin{array}{c} \mathcal{P}_{i}(0)\left[\frac{k_i}{A} \right]^{n} \ \ \ n <  n^* \\ \mathcal{P}_{i}(0) \left[\frac{k_i}{A} \right]^{n} (1-\bar\eta)^{n-n^*}\ \ \ n \ge  n^* \end{array} \right.
\end{equation}
where $\mathcal{P}_i(0)$ is determined from the normalization condition on node $i$ and $A$ is a constant independent of the node that is fixed by the total number of particles $\sum_i n_i = \mathcal{N}$. Their explicit expression is not necessary for  the purposes of the present analysis. 
We can interpret $s = 1/A$ as a chemical potential that fixes the number of particles in the network, and study the "grand-canonical" generating function 
\begin{eqnarray}
\nonumber \mathcal{J}_{N}(s) &= & \sum_\mathcal{N} \sum_{n_1,n_2,\dots, n_N} \delta\left(\sum_i n_i,\mathcal{N}\right) \prod_{i=1}^N \mathcal{P}_i(n_i) \\
& = & \prod_{i=1}^N \left[ \frac{1-(s k_i)^{n^*}}{1-s k_i} + \frac{(s k_i)^{n^*}}{1-(1-\bar\eta) s k_i}\right] .
\end{eqnarray}  
 In the grand-canonical formulation, the particle density can be written as $\nu = \frac{s}{N} \frac{\partial \log \mathcal{J}_{N}(s)}{\partial s}$ \cite{EH05}. We focus now on the limit $n^{*} \to \infty$, for which the particle density becomes  $\nu(s) = \frac{1}{N}\sum_{i} \frac{s k_i}{1- s k_i}$. The sum is defined for values of $s$ smaller than the smallest pole of the argument, i.e. $s < 1/ k_{max}$, i.e. for $A \geq k_{max}$.
 So as long as $A > k_{max}$ the integral converges and the system is able to allocate the corresponding density of particles on the network.
 This is not possible when $A \to k_{max}$, because a pole appears at $k_{max}$. We follow a recent approach by Noh \cite{NSL04} and
  isolate the contribution of the maximum degree nodes $\nu_M(A) = \frac{1}{N} \frac{k_{max}/A}{1-k_{max}/A}$. Grouping together nodes of the same degree, we find in the continuous approximation
\begin{equation}\label{nu}
\nu(A) = \nu_M(A) + \int_{0}^{k_{max}} dk  \frac{P(k) k}{A-k}
\end{equation}
where $P(k) \propto k^{-\gamma}$ is a power-law degree distribution. 
In random networks the maximum degree scales with the system size as $k_{max} \sim \kappa_{max} N^{1/\omega}$, where $\omega$ depends on the generating algorithm (e.g. $\omega = 2,\gamma-1$), it is thus natural to rescale the variables in Eq.\ref{nu}, $A \to a N^{1/\omega}$ and to use $x =k/A$, obtaining
\begin{eqnarray}\label{nu2}
\nonumber \nu(a) & = &\nu_M(a) + N^{\frac{1-\gamma}{\omega}}\int_{k_{min}/A}^{k_{max}/A} dx  \frac{x^{1-\gamma}}{1-x}\\
 & = & \nu_M(a) + \mathcal{O} \left(N^{-\frac{1}{\omega}}\right) - \mathcal{O} \left( N^{\frac{1-\gamma}{\omega}}\right) \log(1-\kappa_{max}/a)  
\end{eqnarray}
where we have estimated the divergence of the integral in the two extremes of integration for $\kappa_{max}/a \to 1$ and $\kappa_{min}/a \to 0$.
Neglecting the second vanishing term and replacing the ratio $\kappa_{max}/a = N \nu_{M}/ (1 + N\nu_{M})$, we get 
\begin{equation}
 \nu(a) \approx   \nu_M(a) + \mathcal{O} (N^{\frac{1-\gamma}{\omega}})\log(1+N\nu_{M}(a)) .
\end{equation}
The last term in the r.h.s. vanishes in the infinite size limit, whereas the first term is finite. This means that  for large systems condensation takes place and a finite fraction of the density is stored in the maximum degree node. 
Decreasing further $A < k_{max}$, the condensation occurs on all nodes of degree $k \geq A$, and the condensate is not localized on a given node. This is an important difference between condensation induced by topological heterogeneity  \cite{NSL04} and standard condensation in homogeneous systems \cite{EH05}.

\begin{figure}
\begin{center}
\includegraphics*[width=0.45\textwidth]{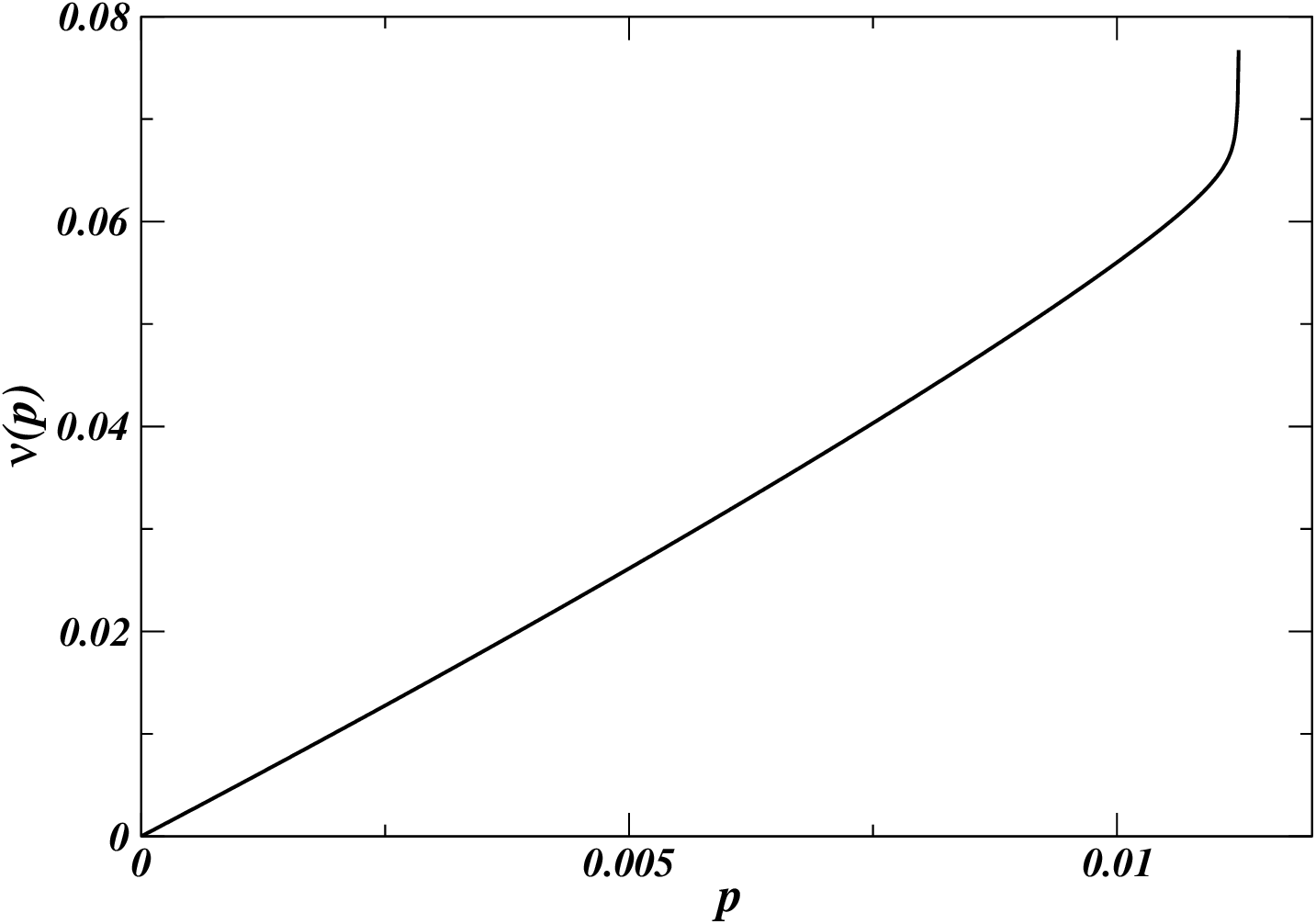}
\caption{ Plot of the function $\nu(p)$ in Eq. \ref{partden}. The divergence occurs at $p=p_c = \mu/[\mu + (1-\mu)k_{max}/z]$. The inset shows that the divergence is triggered by the behavior of $k_{max}$ but nodes of increasing degree contribute to increasingly large portions of the global density. 
} \label{fig_nu}
\end{center}
\end{figure}

\begin{figure}
\begin{center}
\includegraphics*[width=0.45\textwidth]{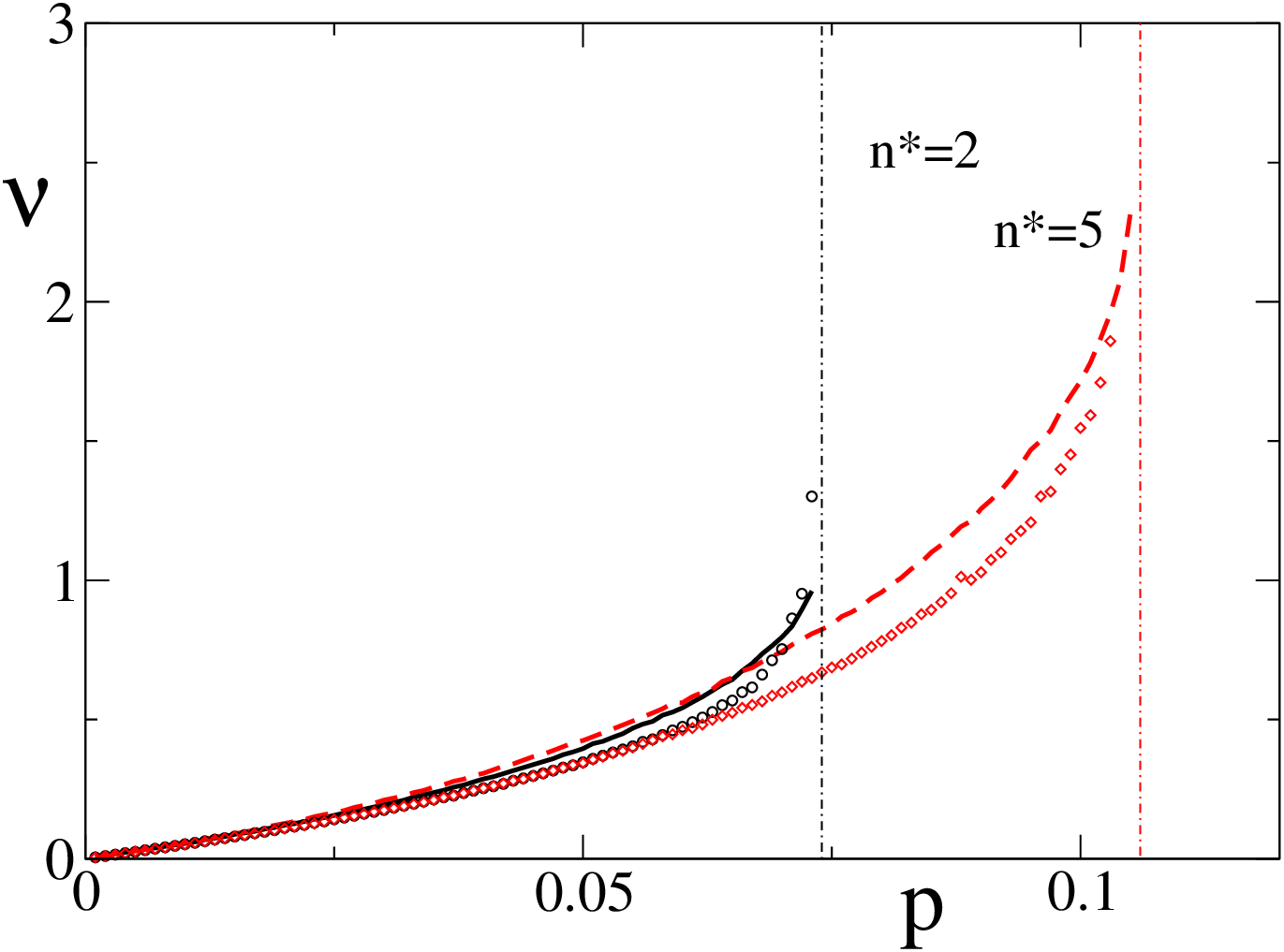}
\caption{ Plot of the density profiles  $\nu(p)$ with ($\bar{\eta}=1$ and several $n*$) and without routing protocol
from numerical simulations run on the same scale-free graph as in Fig. \ref{fig2}.
} \label{fig_nu2}
\end{center}
\end{figure}

The extension to the original non-conserved model is not straightforward, for the absence of a factorized solution, but a qualitative understanding can be  obtained using a mean-field approximation.  
In this way, we can compute the particle density from the generating function (Eq. \ref{genfun})
following the approach of Section \ref{sec-theory2}. 
In the simple case $n^* \to \infty$ the sum can be evaluated analytically, but in this limit the sum diverges as soon as the first fickle nodes appear (because their queues have distribution peaked around $n^*$ that is taken to infinity). The curve exists when $k_F \leq k_{max}$, so $\bar\chi = 0$ and $1-\bar{q} = p/\mu$, thus
\begin{equation}\label{partden}
\nu(p) =  \sum_k P(k) \frac{p+ (1-\mu)\frac{k}{z}\frac{p}{\mu}}{1 - p-(1-\mu)\frac{k}{z}\frac{p}{\mu}} 
\end{equation}
At the critical point $p_c = \frac{\mu}{\mu + (1-\mu)k_{max}/z}$, $\nu(p)$ diverges continuously (see Fig. \ref{fig_nu}). 
The divergence is triggered by nodes of maximum degree, with the same mechanism of condensation, and increasing $p$ above the transition the set of nodes on which particles accumulate grows including nodes of lower and lower degree. 
In Fig.\ref{fig_nu2} we report the behavior of the density of particles as a function of $p$ for systems with finite $n^*$. The data, obtained numerically simulating  the system for $\bar\eta = 0.9$, are compared with the  theoretical prediction, obtained solving the man-field equations for finite $n^* = 2, 5$. The agreement is reasonably good and as expected the density increases for larger values of $n^*$ because more particles can be stored into the queues.

\section{Applications to real traffic systems}
\label{sec-applications}

We have put forward a simple model of non-interacting particles moving randomly and forming queues on networks, and used it to understand the collective behavior that induces congestion phenomena on networks. A possible application is to describe packet-transport in information networks, or vehicular traffic on transportation networks. In the following we briefly discuss these two problems and the properties that can be correctly predicted by simple models inspired to the present one.

\subsection{The Internet at routers level}
\label{sec-applications1}

Congestion in packet-based communication networks means that the network,
or a part of it, is not able to process all arriving information, nodes 
become overloaded, and this implies a global slowing-down of the system.
Congestion phenomena have been observed in wireless networks \cite{HJB04}, in
multimedia networks \cite{CB93}, and, more importantly, in the Internet \cite{J88}.
The first identified Internet's congestion collapse dates
back to October 1986, when data throughput from LBL to
UC in Berkeley suddenly dropped from 32 Kbps to 40 bps.
After that initial event, traffic congestion continued to threaten Internet's practitioners, because of the impossibility of constantly 
monitor and supervise large portions of the Internet, and clearly identify precursors of a congestion event.
For these reasons, understanding congestion phenomena in packet-based communication networks has become a subject of intense 
interdisciplinary research \cite{reg}, with many contributions from 
statistical physicist community,  particularly after the works by Takayasu and collaborators \cite{TT}, 
in which the evidence of a phase transition from a free-flow regime to a congested phase depending on the load level was reported. 

In information networks, like the Internet, packets of information are created at some nodes and then forwarded node-by-node until they reach their destination. 
The packets are dispatched by a routing protocol that tries to minimize the traveling time, taking into account information about the distance and the local traffic. Most of the theoretical models for such a dynamics are based on a shortest-path routing protocol, in which packets follow the shortest-path between a given pair of nodes (where packets are created and destroyed). 
Echenique et al.  \cite{EGM05} have found in
numerical simulations that the nature of the transition
depends on the type of routing rules: in case of purely topological routing (e.g. along the shortest paths),  the congested phase appears continuously, whereas the transition is discontinuous if some traffic-aware scheme is considered (e.g. delivery packets preferentially to uncongested nodes).   
Other works have considered more general forms of routing rules, with particular emphasis on optimization strategies to improve network
performances \cite{TTR04}.

The microscopic dynamics of our model seems quite different from these realistic processes, but it contains all features that are relevant to characterize  the free-flow/congestion transition in information networks, reproducing the collective behaviors already observed in the literature. 
The fact that the absorption of packets occurs only when packets move, not when they are waiting in the queues, mimics the behavior of real packets of information that leave the network when they reach a destination node. On the other hand, random walks and shortest-path routing have very different statistical properties. 
The visiting probability of a node $i$ in the shortest-path routing is proportional to the betweenness centrality of that node, and thus 
 scales non-linearly with its degree, whereas in the random walk protocol the relation is linear. 
In order to accommodate for this statistical feature, and make the routing process more realistic, one could consider degree-biased random walks \cite{FF07}.
Another important ingredient of the model is the
presence of a rejection probability $\eta(n_i)$, that reproduces the {\em congestion avoidance} 
scheme elaborated by computer scientists for the Internet \cite{FJ93}.
This class of algorithms are based on a feedback mechanism that relies on the exchange between 
routers of Acknowledgement signals (ACKs) carrying information on the local level of traffic. 
When the round-trip-time of ACKs sent in a given direction becomes too large, 
the node decreases the rate with which packets are forwarded in such direction. 
Therefore, like in the present model, in the Internet congested nodes have a lower probability to receive packets. 
Such a scheme is useful to retard the onset of the congested phase, but it introduces
a cooperative behavior that can lead to a discontinuous transition.

\begin{figure}
\begin{center}
\includegraphics*[width=0.45\textwidth]{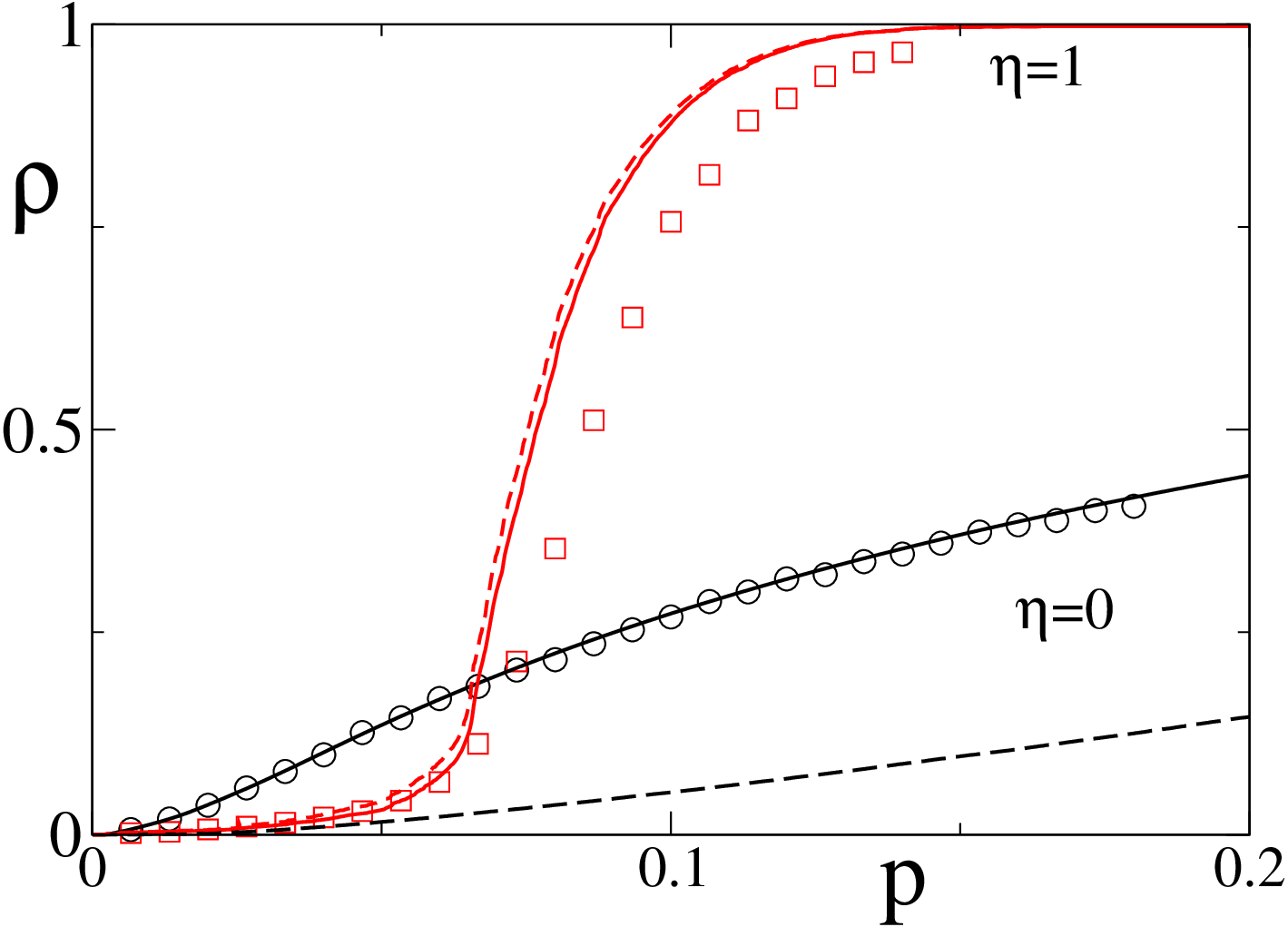}
\caption{Transition curves $\rho(p)$ for the CAIDA network of routers obtained by simulations(points) and theoretical predictions(lines) with $\bar{\eta}=0$
and $\bar{\eta}=1$. We have also reported the corresponding fraction of congested nodes (dotted lines)}
\label{caida-iter}
\end{center}
\end{figure}

In order to characterize the kind of congestion phenomena that could be observed on more realistic topologies than random graphs, we have studied our traffic model on an Internet's map at the routers level obtained by the CAIDA group of Internet's measurements \cite{CAIDA}. The map counts $N = 192244$ nodes, maximum degree $k_{max} = 1071$ and average degree $z=6.3$. It has a degree distribution that is well-fitted by power-law ($P(k) \propto k^{-\gamma}$ with $2 < \gamma < 3$). 
Figure \ref{caida-iter} (left) shows the behavior of the congestion parameter $\rho(p)$ on the routers network obtained running simulations of our traffic model with and without congestion-aware routing protocol. In both cases the congested phase emerges continuously, but with very different behaviors. At low $p$ values the protocol is able to reduce the congestion level, but at larger values of $p$ it is no more effective and the level of congestion starts increasing much faster for $\bar\eta =1$  than for $\bar\eta =0$. 
These results are confirmed by the numerical solution of iterative equations Eqs. \ref{iterative-eq} shown in Fig. \ref{caida-iter}.
In Fig. \ref{caida-iter}, we report also the behavior of the fraction of congested nodes (dotted lines) for both $\bar\eta =0$ and $1$. Comparing the two sets of curves we observe that in the absence of the traffic-aware routing protocol the congested phase is mostly concentrated on a subset of nodes (i.e. the most highly connected ones), whereas for $\bar\eta >0$ the congestion is homogeneously spread over all congested nodes independently of their topological properties.

Another characteristic trait of the Internet is the scaling crossover empirically observed in the single-node's traffic statistics. More precisely it consists in measuring the scaling of traffic's fluctuations $\sigma^2(n)$ with respects to the average traffic on a node $\langle n \rangle$. In Ref. \cite{MGLM08}, it was found that traffic on nodes at the level of the Abilene Internet's sub-network presents two distinct regimes. At low traffic level the fluctuations are linear in the average traffic $\sigma^2(n) \propto \langle n \rangle$, whereas at higher levels of traffic they found $\sigma^2(n) \propto \langle n \rangle^2$ (with prefactors possibly depending on the node's degree). \\
In the present case, for a free node $i$ of degree $k$, the fluctuations of the queue's length are easily obtained within the mean-field approximation (with $n^* \to \infty$).  The generating function gives $G_{k}(s) = 1/(1-a_{k} s)$ with $a_{k} <1$, and  using $G_{k}'(1) = \langle n_{k} \rangle$, we get $\sigma^2_{k} = \langle n_{k} \rangle \left( 1+ \langle  n_{k} \rangle \right)$. At low traffic levels $\langle n_{k} \rangle \ll 1$, thus $\sigma^2_{k} \approx \langle n_{k} \rangle$. Instead when the traffic increases, $\langle n_{k} \rangle$ becomes larger than $1$ and  $\sigma^2_{k} \approx \langle n_{k} \rangle^2$. 

\begin{figure}
\begin{center}
\includegraphics*[width=0.45\textwidth]{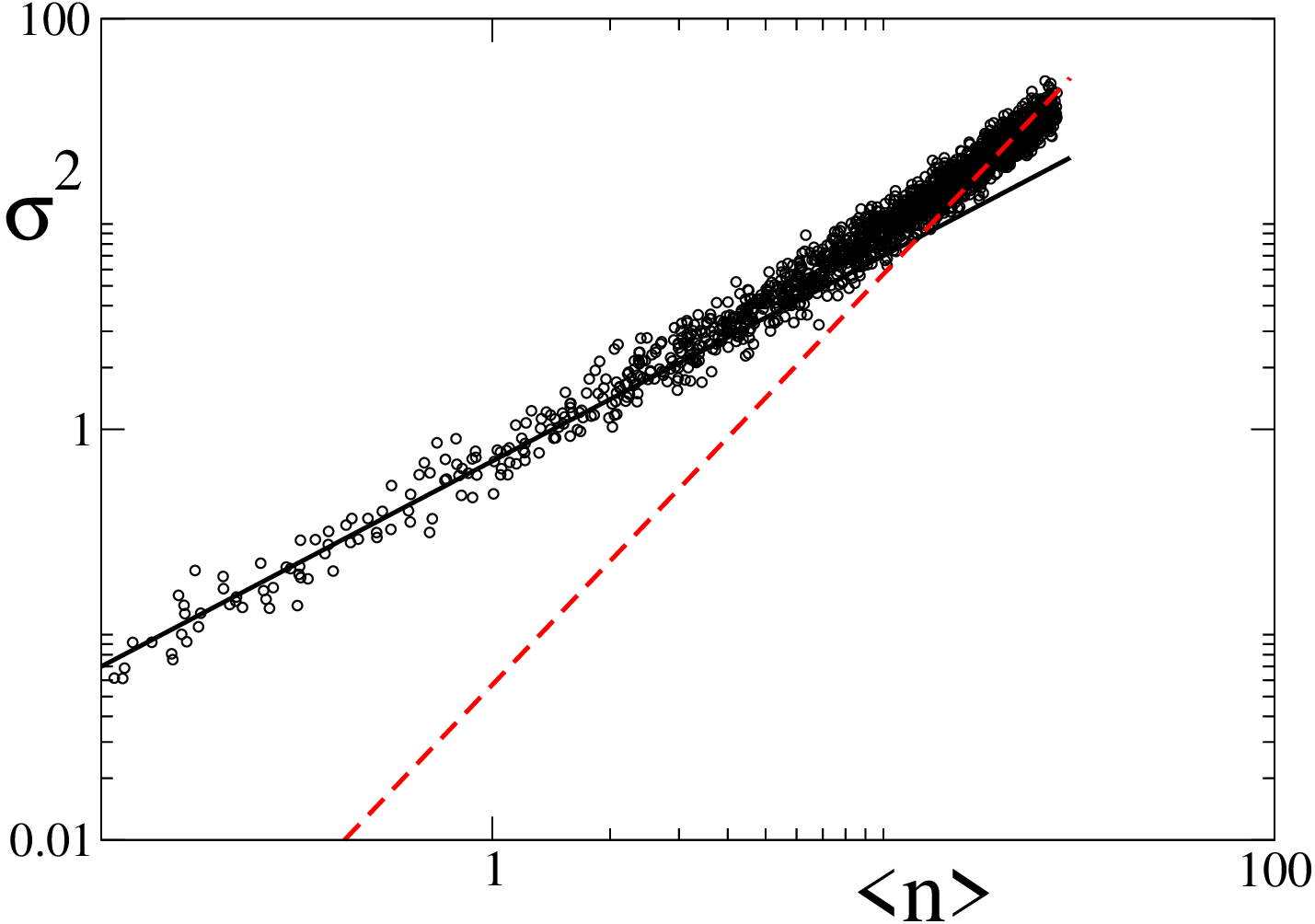}
\caption{Scaling of the fluctuations respect to the mean from simulations onto the CAIDA network.}
\label{scaling}
\end{center}
\end{figure}

We have computed the scaling behavior on the realistic CAIDA's map, using numerical simulations of our model. As expected, the model correctly reproduces this important statistical feature, as evidenced in Fig. \ref{scaling}, in which we display the fluctuation scaling obtained 
averaging over the whole system (i.e. average over the degrees).

\subsection{Traffic in transportation networks}
\label{sec-applications2}
Congestion phenomena are observed in transportation networks as well \cite{H01,KMH05,SFFGL09}, even if the dynamical features can be very different from those defined on information systems.
A first difference is that vehicles flow along roads that are represented by links, whereas the nodes
are just the intersections between them. This is usually accomodated by means of a dual approach, in which nodes and links are exchanged. 
More importantly, road networks have a very peculiar structure with significant constraints imposed by the real space embedding  that influence both the large-scale topology (e.g. minimization of the distance, planar structures), and the single-node properties (road have finite capacity). All these features are expected to have an effect on traffic and thus on the appearance of congestion.

Traffic in urban road networks is usually studied by means of agent-based or fluidynamical models, 
with a large quantity of realistic ingredients \cite{SFFGL09}.
Nevertheless, at a purely statistical level, a simple model of  random walkers with queues can already provide a 
correct qualitative description of the collective behavior of the system. 

\begin{figure}
\begin{center}
\includegraphics*[width=0.45\textwidth]{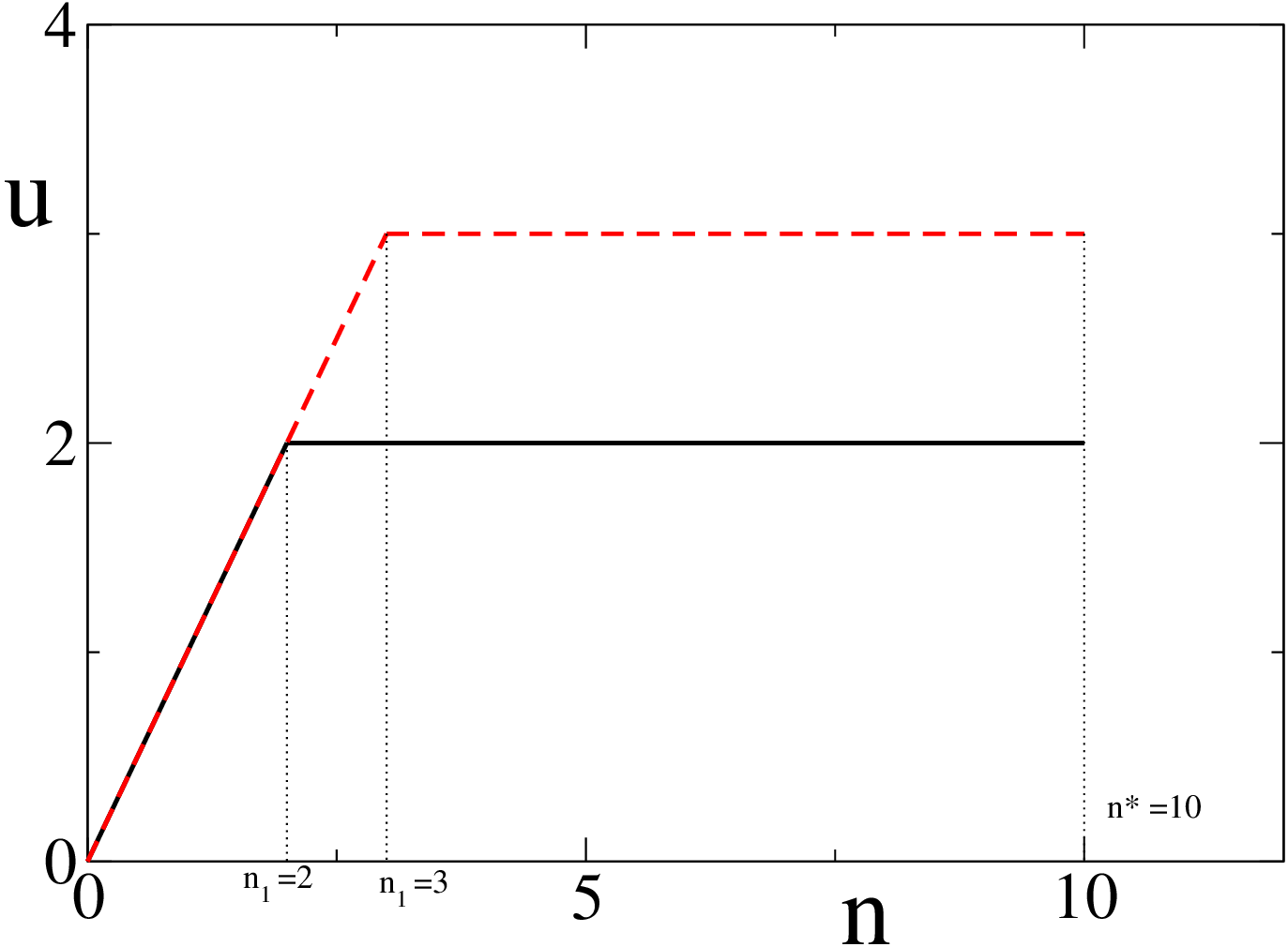}
\caption{
Outgoing flux $u$ from a road (noad) as a function of the number of particles $n$.
Until $n_1$ it is proportional to $n$, and then is constant. The maximum number (capacity) is $n^*$.
If the destination node has $n^*$ particles the flux is zero. 
}
\label{uflux}
\end{center}
\end{figure}

We represent schematically vehicular traffic on a road network considering random walkers on a planar graph. The fact that vehicles proceed along roads in an ordered way is mapped into a queueing process  on the nodes of the dual network. For simplicity we fix the number of particles (or the density $\nu = \mathcal{N}/N$), ignoring 
the agents that may enter or leave the system.  Since roads have a finite capacity,  we fix
the maximum number $n^*$ of particles that a node can store in its queue; and in order to
avoid local overloads, we assume complete rejection when the queue has reached its capacity, i.e. $\eta(n_i)=1$ for $n_i>n^*$. 
Moreover, when the density of cars on a segment of a road is not too high, cars use to travel at a constant speed, that can be mimicked by letting the particles move as non-interacting random walks until a density threshold  $n_1$. At high density, the flux becomes constant because cars form queues. Thus we assume that above $n_1$, particles form queues as well with constant outgoing rate. 
In summary, the flux $u$ out of a road (node) is (see Fig.\ref{uflux}):

\begin{itemize}
\item[{\em i.}] $u(n)=n$ if $n\leq n_1$ and the destination node has less than $n^*$ particles.
\item[{\em ii.}] $u(n)=n_1$ if $n>n_1$ and the destination node has less than $n^*$ particles.
\item[{\em iii.}] $u(n)=0$  if the destination node has $n^*$ particles.
\end{itemize}

Vehicular traffic is usually studied  looking at the {\em density-flux fundamental diagram} \cite{H01}, that is the behavior of the average 
flux $\phi$ as a function of the load of the network, i.e. the density  $\nu$.
The stationary distribution completely factorizes for this system, using the detailed balance. 
We consider  regular lattice and homogeneous rates $r_i = r = 1$ $\forall i$, and the distribution simplifies into:
\begin{itemize}
\item[{\em i.}]  $\mathcal{P}(n) = \mathcal{P}(0)\frac{A^n}{n!}$ if $n\leq n_1$ 
\item[{\em ii.}] $\mathcal{P}(n) = \mathcal{P}(0)\frac{(An_1)^n}{n_1^{n_1} n_1!}$ if $n_1 < n\leq n^*$ 
\item[{\em iii.}] $\mathcal{P}(n) = 0$  if $n>n^*$.
\end{itemize}
From the normalization condition $G(1)=1$ we get $\mathcal{P}(0)$, while $A$ can be obtained 
numerically solving with respect to $A$ the expression for the average density $G'(1) = \nu$.
The average flux $\phi = \langle u \rangle (1-\mathcal{P}(n^*))$ and the average velocity $v=\phi/\nu$ can be easily computed
as function of the parameters of the model.

\begin{figure}
\begin{center}
\includegraphics*[width=0.35\textwidth]{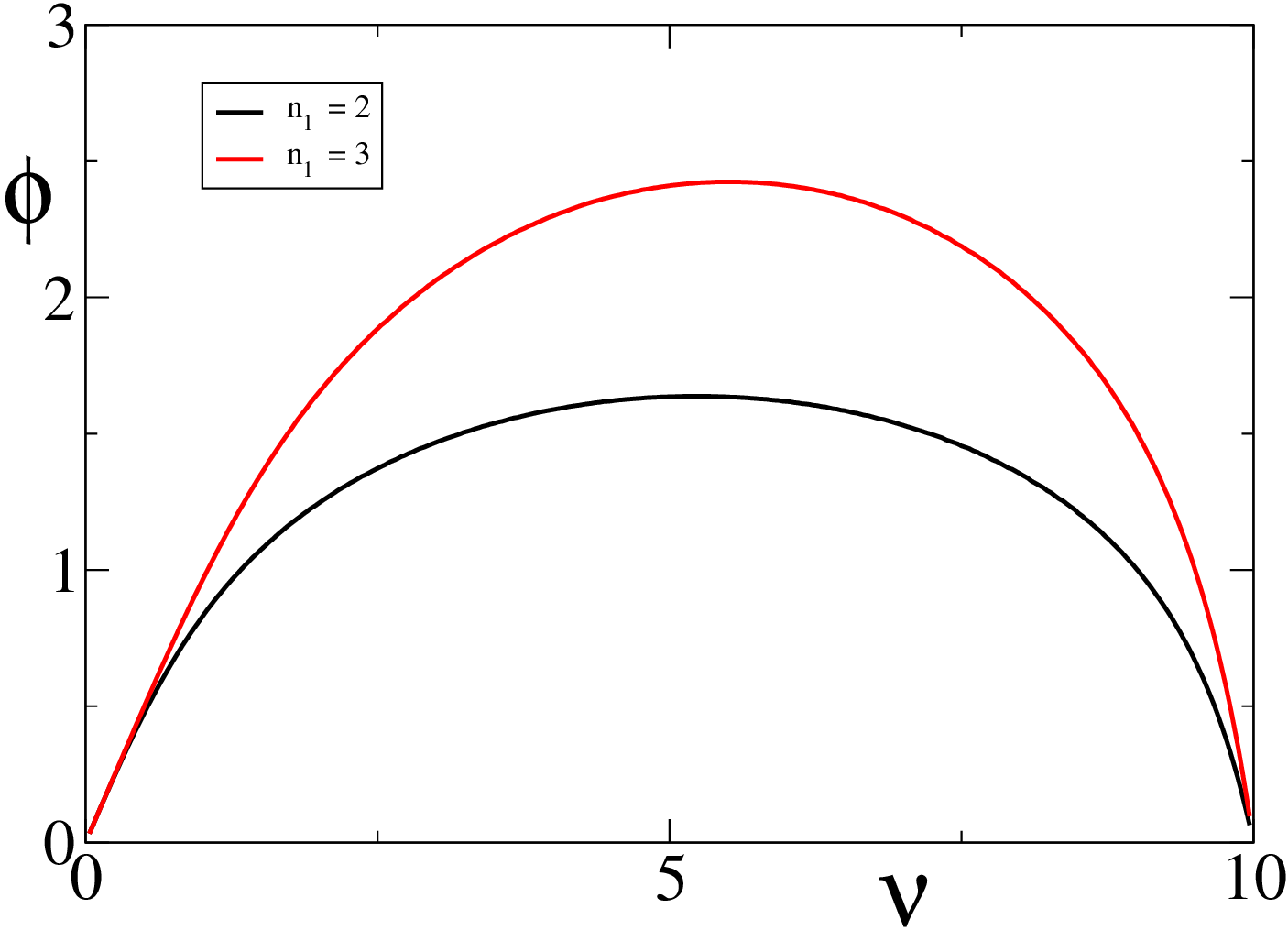}
\includegraphics*[width=0.35\textwidth]{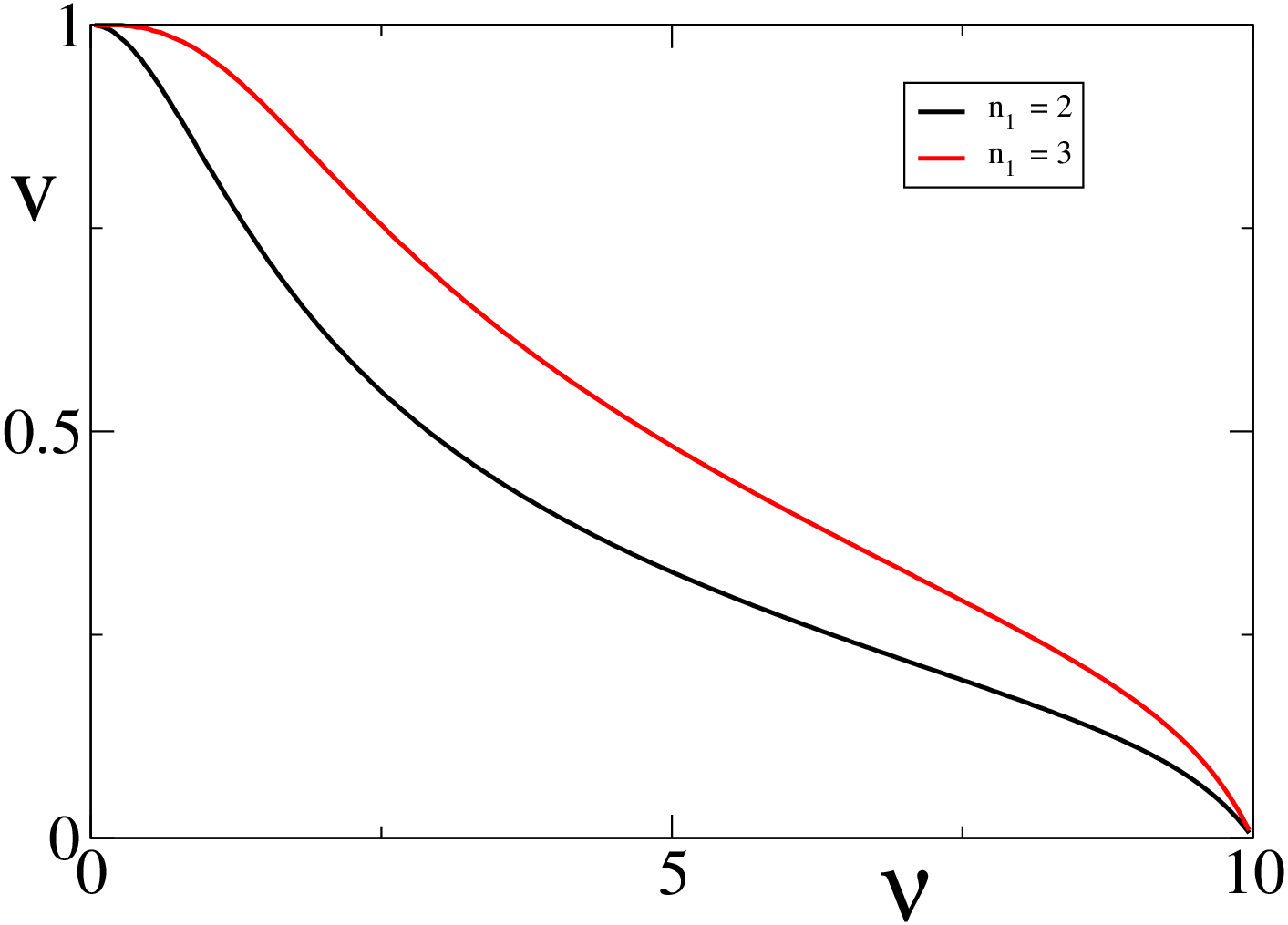}
\caption{
Left: Fundamental diagram for the model onto a regular lattice, 
$n_1=2$ and $3$, $n^*=10$ by analytical calculations.  \\
Right: Velocity profile as a function of the density for the model onto 
a regular lattice $n_1=2$ and $5$, $n^*=10$ by analytical calculations.
}
\label{jam}
\end{center}
\end{figure}

The behavior of the flux $\phi(\nu)$ and the average velocity $v(\nu)$ are reported in Fig. \ref{jam}, for two different
values of $n_1$ and for fixed capacity $n^*=10$.  
Studying these two quantities as a function of the particles density $\nu$, we recover several 
stylized facts of traffic flows in trasportation network \cite{H01}. 
The flow increases almost linearly at low densities
and reaches a maximum  when queues are half-filled. Then the curve decreases until it vanishes at $n^*$. 
Correspondingly, the velocity decreases, from an initial plateau at $v=1$ (free-motion), in a non-linear way (depending on $n^*$ and $n_1$) 
until it vanishes at $\nu = n^*$, where the system becomes completely overloaded.
The congested phase arises always continuously and no phase transition is observed. It would be interesting to understand if a discontinuous transition could be observed in the fundamental diagram by varying some of the parameters of the model.

\section{Conclusions}
\label{sec-conclusions}

We have proposed a minimal traffic model to study congestion phenomena on complex networks, with applications to
realistic processes in information as well as transportation networks.

This traffic model considers particles performing random walks through the network and forming queues on the nodes they visit.
Particles are created on each node with a given rate $p$, but they can be absorbed (at rate $\mu$) only during the hopping process, not when they are waiting into the queues. 
This simple rule is sufficient to generate a transition from a free-flow to a congested phase as a function of the creation/absorption rates.
Moreover, the introduction of a rejection probability $\eta(n)$ for nodes with more than a threshold value $n^*$ of particles can modify the nature of the transition from a continuous to a discontinuous one.

By means of a mean-field approach we have analyzed the behavior of the model on ensembles of generalized random graphs, where the graph properties are specified only by the degree distribution $P(k)$. 
The interplay between the queueing properties of the model and  the topological structure of the network  
generates a rich phenomenology of continuous and discontinuous phase transitions.
In particular, on highly heterogeneous networks, the system can present also an hybrid behavior in which a primary continuous transition to a partially congested state is followed (at higher values of the control parameter $p$) by a discontinuous transition to a state with higher level of congestion (possibly a completely congested state).
For some explicative case we have computed the whole phase diagram with all possible transitions as a function of the parameters of the model. In general on scale free networks the critical line goes to zero in the infinite size limit, but it may converge to a constant if the rejection probability is sufficiently large. 

The present model is reminiscent of a class of non-equilibrium systems (e.g. Zero-Range Processes) in which particles condensation is observed \cite{EH05}. We have shown that a conserved version of our model, without particles' creation and absorption, does present condensation and we have discussed its relation with the congestion phenomenon occurring in the non-conserved model.

We have briefly described two possible applications of the model to packet-based 
traffic in information networks, such as the Internet at the router level, and to the vehicular traffic in road networks.\\
In the case of information networks, the model offers a simplified picture of routing system, in which packets are forwarded
towards destination with a routing protocol that can take into account the level of traffic in the local neighborhood. 
Our results show that traffic-aware routing is useful 
only in heterogeneous networks, where it expands the region of stability of the congestion-free state. 
However, a congested phase may arise abruptly, and then it may persists even under lower traffic loads (hysteresis effects).
The model provides a simple explanation for the so-called "scaling breakdown" of the traffic fluctuations observed on the Internet \cite{MGLM08}, 
in terms of the intrinsic statistical property of the queues.\\
On the other hand a slightly modified version of the conserved model, in which nodes have a maximum capacity,
can reproduce qualitatively the form of the fundamental diagram and velocity profile observed for the vehicular traffic in transportation networks \cite{H01}. 

These results show that a simple model of interacting random walks with some further ingredient is enough to obtain a very good description of the statistical properties of the collective behavior of the system.

%The mechanisms triggering the emergence of congestion is somewhat reminiscent of  jamming or bootstrap percolation,
%where a node is occupied if the number of occupied neighbors exceeds a given threshold. Also in these models, 
%as the threshold increases, the transition turns from continuous to discontinuous \cite{DGM06}.

The work presented here can be extended in several interesting directions. 
%The analogy with models of urban traffic is also a promising avenue. In the dual representation of the road network 
%\cite{Latora}, nodes represent segments of roads and links denote junctions.  
%A model in which adaptive drivers on a road network try to avoid congested roads exhibits a similar phenomenology 
%\cite{urban}: as the level of traffic increases, drivers find ways to avoid overloaded streets, 
%thus distributing as uniformly as possible the traffic load. However, at a critical threshold a congestion 
%phase transition takes place beyond which the system is plagued by strong traffic fluctuations. 
%Similarly, in the present model, traffic-aware routing makes the traffic load uniform on a large part of the network.
%The occurrence  of a discontinuous phase transition can indeed be traced back to this homogenizing effect: 
%when traffic control is strong, a finite fraction of the network can suddenly become congested upon 
%increasing the traffic load ($p$). 
The possibility to solve the model on a given network (e.g. Internet's map or urban road networks) using Eqs. \ref{iterative-eq} with realistic parameters, could provide both specific predictions on the robustness of the network to traffic overloads and important hints 
for the design of systems that could be less vulnerable to congestion phenomena. The dynamical environment created within this model could also be exploited as a framework for testing the statistical properties of single particle dynamics under more complex routing schemes, in a way similar to the study of a tracing-particle in hydrodynamics. 

Finally, it would be interesting to model the complex adaptive behavior of human users in 
communication networks, such as the Internet, by introducing variable rates of packets production in response to network performances. It is known that users face the social dilemma of maximizing their own communiction rates, 
maintaining the system far from the congested state \cite{HL97}. 
In such a situation, the presence of a continuous transition may allow the system to self-organize 
at the edge of criticality, whereas a discontinuous transition may have catastrophic consequences.

\appendix
\section{Validity of Product-measure distributions}
\label{app1}
The existence of factorized stationary states in interacting particles systems on graphs is a subject of great interest, because models with this property are exactly solvable and can be used to investigate complex dynamical phenomena like mass-transport or condensation. 
The most studied family of models with a product-measure stationary state are the zero-range processes (ZRP), in which it comes directly from the fact that hop rates depend only on the number of particles in the departure node \cite{EH05}. 
In the Misanthrope processes, instead, the hop rate depends on both departure and arrival node, but under some conditions the stationary state distribution is still factorizable on single nodes \cite{EH05}.
More general sufficient conditions to have factorized stationary states in continuous mass-transport models on general graphs were pointed out recently by Evans et al. \cite{EMZ04}. 

All statistical physics models on which factorization has been proved have either a conserved number of particles or zero-range interaction,
and satisfy either a detailed balance condition or a pairwise balance condition, depending on whether the transition rates are symmetric or not along the links of the network. A partial generalization of these results can be obtained applying  Jackson's Theorem, a well-known result in queueing theory, that requires only the stationarity condition and a local conservation law for the fluxes of particles \cite{BGDT06}.
To our knowledge there is no general result for factorization in models with non-conserved number of particles in which the hop rates depends on the number of particles in both departure and arrival nodes. 
The aim of this appendix is to discuss in detail some special cases in which factorization is exact. 

\subsection{Detailed Balance Condition}
\label{app1-1}
Let us consider the conserved model in which $\mathcal{N}$ particles are deployed on a general graph and can hop from node $i$ to $j$ with hop rate $u_{ij}(n_i, n_j)$, i.e. a general Misanthrope process. In our model, we will specify later $u_{ij}(n_i,n_j) = r_i W_{ij} [1-\eta_j(n_j)]$ with $W_{ij} = \frac{1}{k_i}$.  
When there are no currents in the graph, the detailed balance should be valid, and this is the case of models in which the hop rate is symmetric along the edges of a node. The dynamics is based on random walks, thus we expect detailed balance to hold.

Given the states $\underline{n} = (n_1,\dots, n_i,\dots, n_j,\dots, n_N)$ and $\underline{n}' = (n_1,\dots, n_i+1,\dots, n_j -1,\dots, n_N)$, with transition rates $w(\underline{n} \to \underline{n}') = u_{ji}(n_j,n_i)$ and $w(\underline{n}' \to \underline{n}) = u_{ij}(n_i+1,n_j-1)$,
imposing the detailed balance condition and introducing the factorized ansatz $\mathcal{P}(n_1, \dots, n_N) = \prod_i \mathcal{P}_i(n_i)$, 
we get the equation
\begin{equation}\label{app_balance1}
u_{ij}(n_i+1,n_j-1) \mathcal{P}_i(n_i+1) \mathcal{P}_j(n_j-1)  = u_{ji}(n_j,n_i) \mathcal{P}_i(n_i) \mathcal{P}_j(n_j) 
\end{equation}
The case $n_i = 0$, $n_j >0$ gives $u_{ij}(1,n_j-1) \mathcal{P}_i(1) \mathcal{P}_j(n_j-1) = u_{ji}(n_j,0) \mathcal{P}_i(0) \mathcal{P}_j(n_j)$, that is a recurrence equation producing
\begin{equation}\label{dist_0}
\mathcal{P}_j(n_j) = \mathcal{P}_j(0) \left[\frac{\mathcal{P}_i(1)}{\mathcal{P}_i(0)}\right]^{n_j} \prod_{\ell =1}^{n_j} \frac{u_{ij}(1,\ell-1)}{u_{ji}(\ell,0)}
\end{equation}
For this result to be correct also for $n_i > 0$, we plug it into Eq. \ref{app_balance1} and obtain a condition on the hop rates
\begin{equation}\label{rate_cond}
u_{ij}(n_i+1,n_j-1) \frac{u_{ij}(1,0)}{u_{ji}(1,0)} \frac{u_{ji}(1,n_i) u_{ji}(n_j,0)}{u_{ij}(n_i+1,0) u_{ij}(1,n_j-1)}   = u_{ji}(n_j,n_i) 
\end{equation}

The condition is satisfied by the hop rates $u_{ij}(n_i,n_j) = \frac{r_i}{k_i}[1-\eta_j(n_j)]$ of our conserved model, that thus admits 
a product-measure stationary distribution. Reasonably assuming $\eta(0) = 0$ and using properties from Eq. \ref{dist_0}, the single-site distribution becomes 
\begin{equation}\label{fi0}
\mathcal{P}_i(n) = \mathcal{P}_i(0) \left(\frac{k_i}{A r_i}\right)^n \prod_{\ell =1}^{n} \left[1-\eta_i(\ell-1)\right]
\end{equation}
where $\mathcal{P}_i(0)$ is fixed by the normalization and $A$ is a constant independent of the node. 

When we introduce particle creation and destruction, detailed balance is not always satisfied on general graphs. An example is provided by  the non-conserved model without rejection probability. Suppose the particles are created on node $i$ with rate $p_i$,  and  absorbed when they move to $j$ with probability $\mu_j$. Note that the particles transfer occurs now with rate $u_{ij}(n_i,n_j) = \frac{r_i}{k_i}(1-\mu_j)$.
Writing the detailed balance for the particle-transfer transition and factorizing the distribution we find the relation $\frac{r_i}{k_i}(1-\mu_j) \mathcal{P}_i(n_i+1) \mathcal{P}_j(n_j-1) = \frac{r_j}{k_j}(1-\mu_i) \mathcal{P}_i(n_i) \mathcal{P}_j(n_j)$ that gives 
\begin{equation}\label{dist_1a}
\mathcal{P}_j(n_j) = \mathcal{P}_j(0) \left[\frac{r_i k_j \mathcal{P}_i(1)}{r_j k_i \mathcal{P}_i(0)}\right]^{n_j} \left[\frac{1-\mu_j}{1-\mu_i}\right]^{n_j} 
\end{equation}
i.e.
\begin{equation}\label{dist_1b}
\mathcal{P}_i(n) = \mathcal{P}_i(0) \left[\frac{k_i }{r_i A}\right]^{n} (1-\mu_i)^{n}
\end{equation}
where $A$ is again a constant independent of the node properties. This expression satisfies also the condition for general $n_i > 0$, but when plugged it into the  detailed balance condition for the creation-absorption transition, i.e. $p_i \mathcal{P}_i(n_i) = \frac{r_i}{k_i} \sum_{j\in v(i)} \mu_j \mathcal{P}_i(n_i+1)$, it produces a node-dependent expression for $A$. Only in the special case of 
 an homogeneous set of parameters on a regular graph (i.e. $k_i = k \ \forall i$), detailed balance is enough to have a product-measure, that reads $\mathcal{P}(n) = (1- p/r \mu) (p / r \mu )^{n}$.

Hence, detailed balance is not a sufficient condition for factorization of our model on a general graph, not even in the simpler case in which particles are not rejected by destination nodes. In the next section, we will check if a product-measure can be derived from the weaker condition imposed by stationarity.  

\subsection{Stationarity condition and the Jackson's Theorem}
\label{app1-2}
The reason of the failure of detailed balance is not the particular mechanism of particles destruction that we have considered.
Indeed, a non-conserved model with standard birth-death process at the nodes (i.e. particles are created with rate $p_i$ and destroyed with rate $\mu_i$) would have the same problems. Nonetheless, such a model admits a product-measure stationary distribution. This can be easily proved using Jackson's Theorem, one of the major results in queueing theory \cite{BGDT06}. 

Jackson's Theorem is usually formulated for networks of queues with exponentially distributed service times and Poisson arrivals from outside
and for this reason it can be easily adapted to describe the present system. 
The first ingredient of Jackson's approach is the global balance, or stationarity condition, of the probability distribution, 
\begin{eqnarray}\label{app_stationary}
 \left(\sum_{i=1}^N p_i + \sum_{i=1}^N r_i \sum_{j = 1}^{N} W_{ij} [1-\eta_j(n_j)]\right) \mathcal{P}(n_1, \dots, n_N)   = \\ \nonumber \sum_{i=1}^N p_i (1-\delta_{n_i,0}) \mathcal{P}(n_1, \dots, n_i-1, \dots, n_N)  \\  \nonumber + \sum_{i=1}^N r_i \sum_{j=1}^N \mu_j W_{ij} [1-\eta_j(n_j)] \mathcal{P}(n_1, \dots, n_i+1, \dots, n_j, \dots, n_N)  \\ \nonumber +  \sum_{i=1}^N \sum_{j=1}^N r_j W_{ji} (1-\delta_{n_i,0})(1-\mu_i) [1-\eta_i(n_i-1)] \mathcal{P}(n_1, \dots, n_i-1, \dots, n_j+1, \dots, n_N) 
\end{eqnarray}
with $W_{ij} = 1/k_i$.
We  focus here on the dynamics without rejection ($\eta(n) = 0$ $\forall n$) and consider two important additional relations.
In the stationary un-congested state, the incoming flux of particles in the network has to be balanced by the outcoming flux. Let us call 
$\lambda_i$ the stationary rate of particles entering node $i$, then the balance of particles entering and leaving the network is given by 
\begin{equation}\label{global-cond}
\sum_i p_i = \sum_i \lambda_i \sum_j W_{ij} \mu_j .
\end{equation} 
The rates $\{ \lambda_i\}$ are defined by means of a local conservation law at the nodes, 
\begin{equation}\label{local-cond}
\lambda_i = p_i +  (1-\mu_i) \sum_j \lambda_j W_{ji} .
\end{equation}
If a solution exists, this system of linear equations can be solved to find $\{ \lambda_i\}$ for any given graph and set of parameters $\{p_i, \mu_i\}$.

Once we have computed $\{\lambda_i\}$, the stationary probability distribution (in the uncongested phase) can be expressed in the following product-measure form,
\begin{equation}\label{app_JT}
\mathcal{P}(n_1, \dots, n_N) =  \prod_{i=1}^{N} \mathcal{P}_i(n_i)  =  \prod_i \left( 1 - \frac{\lambda_i}{r_i}\right) \left( \frac{\lambda_i}{r_i} \right)^{n_i}
\end{equation}
The proof of factorization derives straightforwardly from the stationarity condition. Inserting expression  \ref{app_JT} into Eq. \ref{app_stationary} we get 
\begin{eqnarray}
\nonumber \sum_i p_i + \sum_i r_i & = \sum_i p_i \frac{\mathcal{P}_i(n_i-1)}{\mathcal{P}_i(n_i)} + \sum_{ij} r_i W_{ij} \mu_j \frac{\mathcal{P}_i(n_i+1)}{\mathcal{P}_i(n_i)} \\ & \quad + \sum_{ij} r_j W_{ji} (1-\mu_i) \frac{\mathcal{P}_i(n_i-1)}{\mathcal{P}_i(n_i)}\frac{\mathcal{P}_j(n_j+1)}{\mathcal{P}_j(n_j)} 
\end{eqnarray} 
i.e. 
\begin{eqnarray}
\nonumber \sum_i \left(p_i +  r_i \right) & = \sum_i \frac{p_i r_i}{\lambda_i}  + \sum_{ij} r_i W_{ij} \mu_j \frac{\lambda_i}{r_i}  + \sum_{ij} r_j W_{ji} (1-\mu_i) \frac{r_i \lambda_j}{r_j \lambda_i} \\
\nonumber & = \sum_i \frac{p_i r_i}{\lambda_i} + \sum_i p_i + \sum_{i} \frac{r_i}{\lambda_i} \left( \sum_j (1-\mu_i) \lambda_j W_{ji} \right)  \\
& = \sum_i \frac{p_i r_i}{\lambda_i} + \sum_i p_i + \sum_i \frac{r_i}{\lambda_i} (\lambda_i - p_i) = \sum_i \left( p_i + r_i \right)
\end{eqnarray} 
where we have used Eqs. \ref{global-cond} and \ref{local-cond}.

\section{Pseudo-factorization in the general non-conserved model}
\label{app2}
In this Appendix, we show  that the probability distribution of the general model proposed in this paper is not exactly factorizable over the nodes, i.e. a product-measure stationary state  does not exist. Nevertheless, the way in which factorization is broken is very peculiar and still allows one to obtain important information on the structure of the stationary distribution. 

\subsection{Beyond Jackson's Theorem}
\label{app2-1}
If we consider also rejection of particles at the nodes, Jackson's Theorem does not apply, though we can prove a theorem that specifies the form of the stationary distribution.
\begin{thm}\label{th1}
Consider an open particles system on a general graph with adjacency matrix $\{ a_{ij} \}$ in which particles enter the system at every node $i$ with rate $p_i$, hop from node $i$ to node $j$ with rate  $u_{ij}(n_i,n_j) = r_i  W_{ij} (1-\mu_j) [1- \eta_j(n_j)]$ (with $W_{ij} = a_{ij}/k_i$), and leave the system with rate $u_{i0} = r_i \sum_j W_{ij} \mu_j [1-\eta_j(n_j)]$.  Let $\{ \lambda_i (n_i,\underline{n}_{-i}) \}$ be functions of the configuration $(n_i, \underline{n}_{-i}) = \underline{n} = (n_1, \dots, n_i, \dots, n_N)$ satisfying the system of linear equations 
\begin{equation}\label{lambdan}
\lambda_i(n_i,\underline{n}_{-i}) \sum_j W_{ij} [1-\eta_j(n_j-1)] = p_i + \sum_j \lambda_j(n_j,\underline{n}_{-j}) W_{ji} (1-\mu_i) [1-\eta_i(n_i-1)]  
\end{equation}
then the stationary probability distribution of particles is 
\begin{equation}\label{app_JTbeyond}
\mathcal{P}(n_1, \dots, n_N) \propto  \prod_{i=1}^{N} \mathcal{P}_i(\underline{n})  =  \prod_{i=1}^{N} \prod_{\ell=1}^{n_i} \frac{\lambda_i(\ell-1,\underline{n}_{-i})}{r_i} 
\end{equation}
\end{thm}

The proof of the theorem is straightforward and consists in inserting the factorized form in the stationarity condition like in Jackson's Theorem, 
\begin{eqnarray}
\nonumber \sum_i p_i + \sum_i r_i \sum_j W_{ij} [1-\eta_j(n_j)] = \\ 
\quad \sum_i p_i \frac{\mathcal{P}_i(n_i-1)}{\mathcal{P}_i(n_i)} + \sum_{ij} r_i W_{ij} \mu_j [1-\eta_j(n_j)] \frac{\mathcal{P}_i(n_i+1)}{\mathcal{P}_i(n_i)} \\  
\quad + \sum_{ij} r_j W_{ji} (1-\mu_i) [1-\eta_i(n_i-1)]\frac{\mathcal{P}_i(n_i-1)}{\mathcal{P}_i(n_i)}\frac{\mathcal{P}_j(n_j+1)}{\mathcal{P}_j(n_j)} 
\end{eqnarray} 
i.e. the r.h.s. reads
\begin{eqnarray}
\nonumber  = \sum_i \frac{p_i r_i}{\lambda(n_i,\underline{n}_{-i})} + \sum_{ij} r_i  W_{ij} \mu_j [1-\eta_j(n_j)] \frac{\lambda_i(n_i+1,\underline{n}_{-i})}{r_i} \\
\nonumber \quad + \sum_{ij} r_j W_{ji} (1-\mu_i) [1-\eta_i(n_i-1)] \frac{\lambda_j(n_j+1,\underline{n}_{-j}) r_i}{\lambda_i(n_i,\underline{n}_{-i}) r_j} \\
\nonumber  =   \sum_i \frac{ r_i}{\lambda(n_i,\underline{n}_{-i})} \left[ p_i + [1-\eta_i (n_i-1)](1-\mu_i) \sum_jW_{ji} \lambda_j(n_j+1,\underline{n}_{-j}) \right]  \\
\nonumber \quad + \sum_{i} \lambda_i(n_i+1,\underline{n}_{-i}) \sum_j  \mu_j W_{ij} [1-\eta_j(n_j)]  \\
\nonumber  = \sum_i \frac{ r_i}{\lambda(n_i,\underline{n}_{-i})} \left[\lambda_i(n_i,\underline{n}_{-i}) \sum_j W_{ij} [1-\eta_j(n_j)] \right] \\
\nonumber \quad + \sum_{i} \lambda_i(n_i+1,\underline{n}_{-i}) \sum_j  \mu_j W_{ij} [1-\eta_j(n_j)]  \\
 = \sum_{ij}  r_i  W_{ij} [1-\eta_j(n_j)]  + \sum_{i} \lambda_i(n_i+1,\underline{n}_{-i}) \sum_j  \mu_j W_{ij} [1-\eta_j(n_j)]  
\end{eqnarray} 
where in the last line we have used the local flux-balance condition (\ref{lambdan}).
The stationarity condition thus becomes a global balance condition $\sum_i p_i = \sum_i \lambda_i(n_i+1,\underline{n}_{-i}) \sum_j \mu_j W_{ij} [1-\eta_j(n_j)]$, that is verified by summing over all nodes the local balance conditions in Eq.\ref{lambdan}.

Note that the theorem holds also for  more general hop rates $u_{ij}(n_i,n_j)$  and for  creation/destruction of particles that also depend on the number of particles in neighboring nodes.

\subsection{Derivation of mean-field equations and role of correlations}
\label{app2-2}
The mean-field approximation consists in studying the single-node behavior, neglecting the correlations that could exist between different nodes. The pseudo-factorized form of the stationary distribution suggested by Theorem \ref{th1} says that correlations only exist  at the level of the equations defining the set of $\{\lambda_i \}$. These self-consistent equations could be solved in principle for each configuration $\underline{n}$, but this is impossible in practice and we have to resort to some approximation. 

Averaging over $n_i,n_j$ both sides of Eq. \ref{lambdan}, and replacing the rejection function $\eta(n)$ with the corresponding rejection probability $ \chi$,  leads to the mean-field equation 
\begin{equation}\label{lambdan2}
\frac{\langle \lambda_i \rangle}{k_i} \sum_{j \in v(i)} [1-\chi_j] = p_i + \sum_{j \in v(i)} \frac{\langle \lambda_j \rangle}{k_j} (1-\mu_i) [1- \chi_i]  
\end{equation}
that is valid up to the congestion transition.

Note that the quantity $\lambda_i$ represents the total rate of particles outcoming from node $i$.  As particles hop with rate $r_i$ per node, independent from the number of particles $n_i$, we expect $\lambda_i$ to be proportional
to  $r_i$ and to the probability that the queue in $i$ is not empty, that we call $1-q_i$. 
With $\langle \lambda_i \rangle \approx r_i (1-q_i)$, the Eq. \ref{lambdan2} becomes identical to the condition $\langle \dot{n}\rangle = 0$
used in Section \ref{sec-theory1} to derive a system of $2N$ mean-field equations for the node quantities $\{q_i, \chi_i \}$ that can  be solved recursively on every graph. 

\begin{figure}
\begin{center}
\includegraphics*[width=0.4\textwidth]{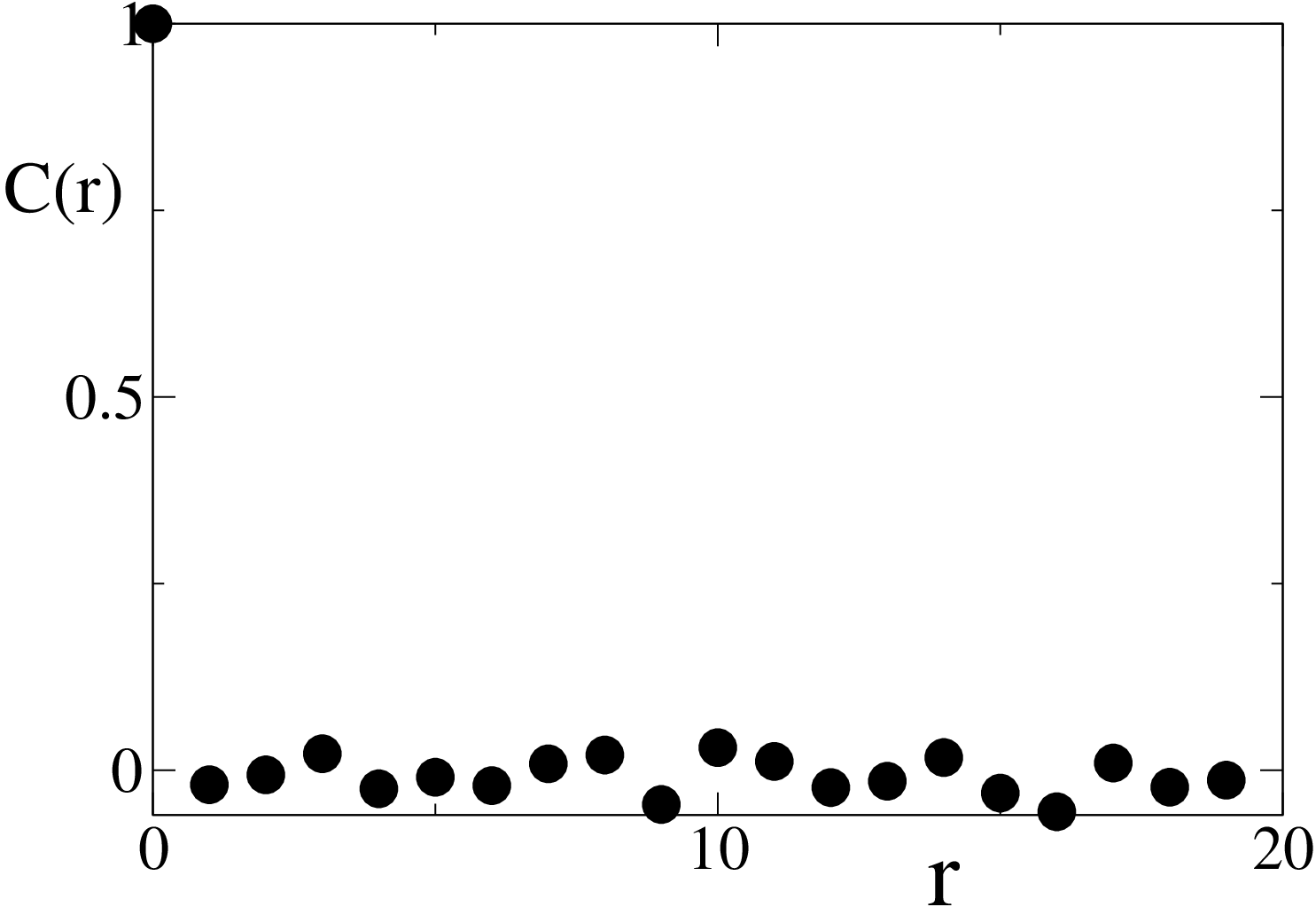}
\caption{
Normalized correlation in the number of packets between a node and a node placed at distance $r$ 
in a square lattice with periodic boundary conditions of size $L=500$, $p=0.12$, $\mu=0.2$ and $\bar\eta = 0.5$, $n^*=10$.
}
\label{correl}
\end{center}
\end{figure}

When all nodes are far from a congested state, factorization is almost exact, because the rejection term is effectively inactive. We have verified this statement computing the two-points correlations $C(r) \equiv \langle n_i n_j\rangle$ ($\Vert i -j \Vert = r$) between the queue lengths of nodes at distance $r$ on a regular square lattice (see Fig. \ref{correl}). For  $p < p_c$, correlations are completely absent.

\section{Behavior of the system for small queueing capacity}
\label{app3}

\begin{figure}
\begin{center}
\includegraphics*[width=0.35\textwidth]{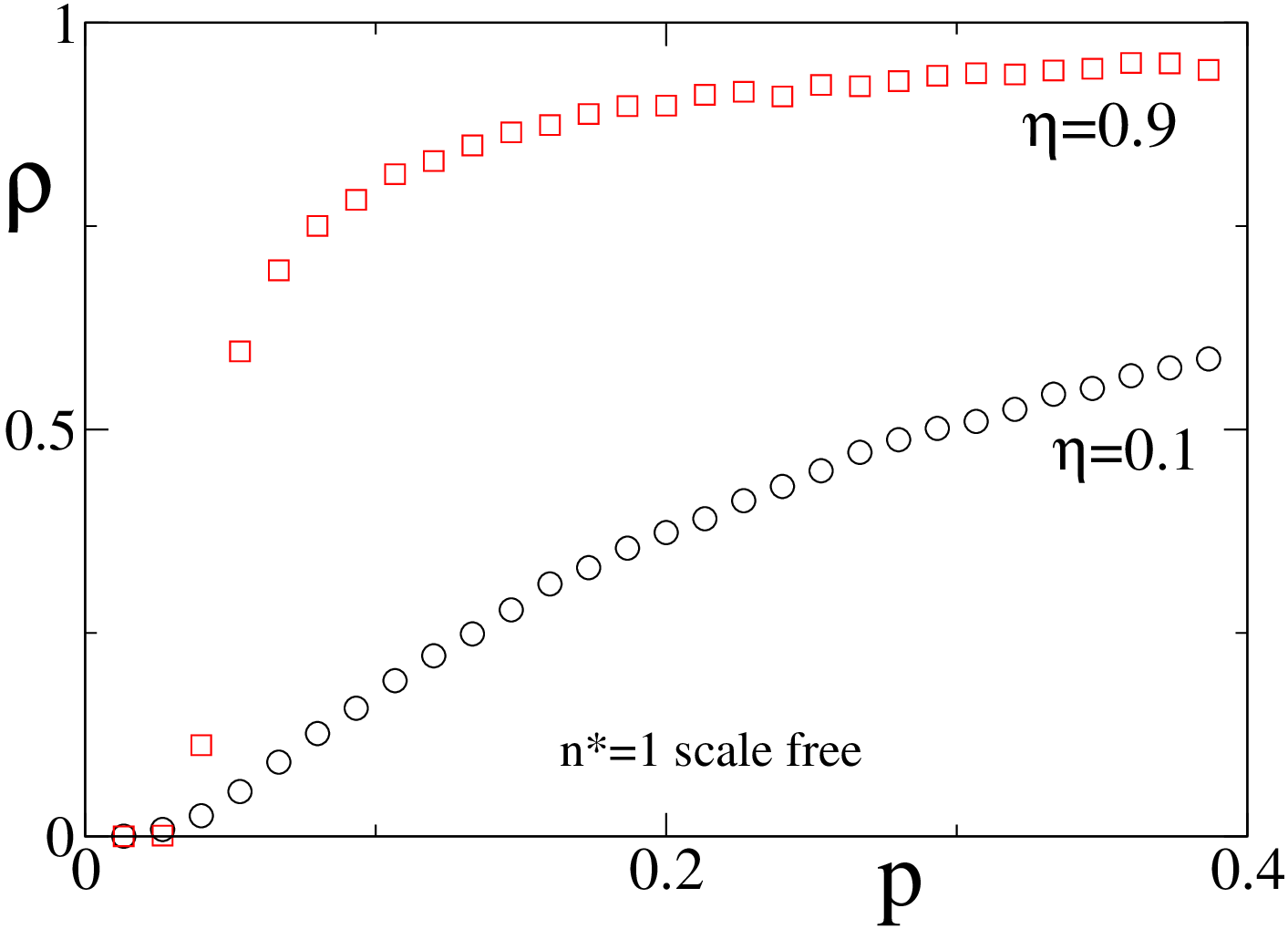}
\includegraphics*[width=0.35\textwidth]{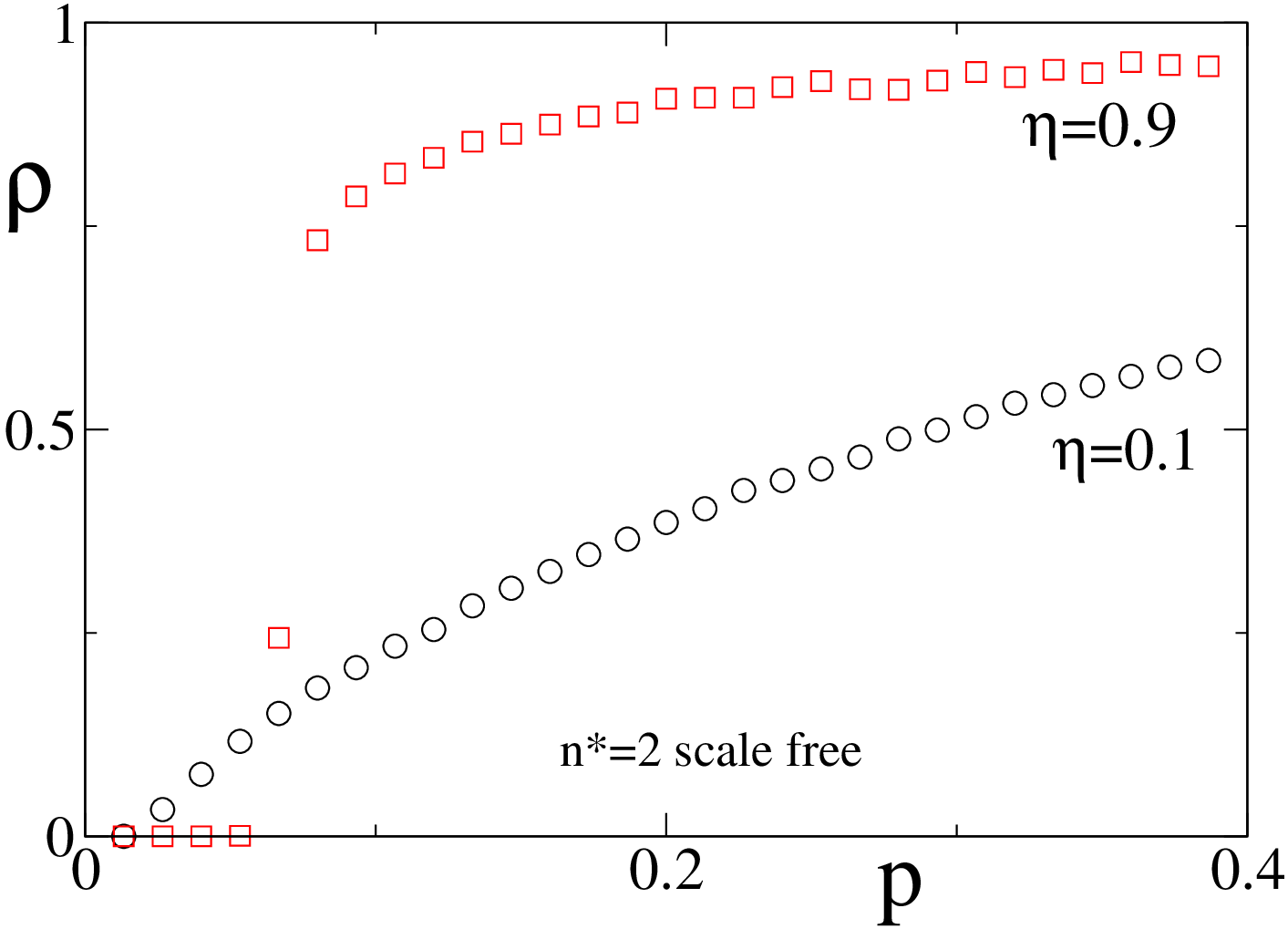}
\caption{Behavior of the congestion parameter $\rho(p)$ on scale-free network ($N=3000$, $\gamma=3$) for $\bar\eta = 0.1$ (black circles), and  $0.9$ (red squares) with $n^* = 1$ (left) and $n^* = 2$ (right).
}
\label{ncut}
\end{center}
\end{figure}

Most of the analytical results presented in the main text, both at the ensemble level and on single graphs, are obtained 
in the limit $n^* \to \infty$.
For finite $n^*$, the calculations still can be solved numerically but the overall approach becomes cumbersome and much less transparent.
We have stressed that, except for the pathological case of the average density $\nu(p)$ (see Sec. \ref{sec-theory3}), the qualitative behavior of the system does not change if we consider finite but large $n^*$. 
It is natural to ask how far we can push this approximation, and if the behavior for small values of $n^*$ is still qualitatively similar to that for $n^* \to \infty$.  For $n^* = 1$, i.e. when each node reject particles with a probability $\bar\eta$ as soon as it contains a particle, 
mean-field calculations on ensembles of random graphs with a given degree distribution do not present further difficulties compared to the $n^* \to \infty$ case. Fig.\ref{ncut} (left) reports the behavior of the congestion order parameter $\rho(p)$ on a scale-free network with rejection probability $\eta(n) = \bar\eta \theta(n-1)$ for $\bar\eta = 0.1$ (black circles) and $\bar\eta = 0.9$ (red squares). 
Fig.\ref{ncut} (left) shows that increasing $\bar\eta$ a discontinuous transition to the congested phase appears, but without any shift of the transition point to higher values of $p$. Hence, in this case, the traffic-aware protocol is not effective in enlarging the free phase.
Nevertheless, for $n^* =2$ (right panel in Fig. \ref{ncut}) the behavior is already qualitatively the same that in the limit $n^* \to \infty$. 

\section{Metastability of the free phase}
\begin{figure}
\begin{center}
\includegraphics*[width=0.45\textwidth]{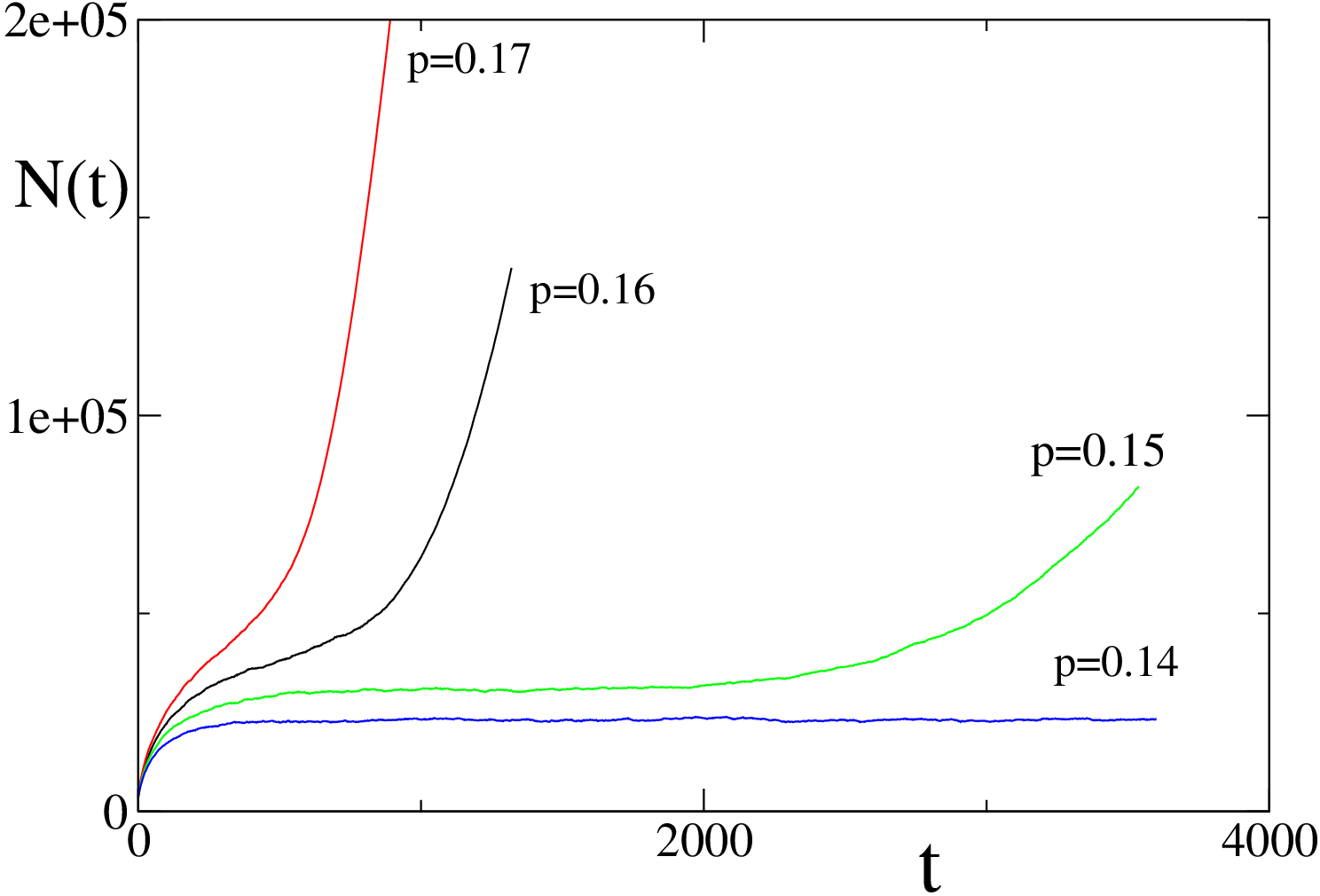}
\caption{Time series of the total number of packets from simulations onto a square lattice $100X100$ with pbc, $\mu=0.2$, $\eta=0.9$, $n^*=10$
starting from the free phase, for different values of $p$. At $p=0.15$, after a transient, the system jumps
into the congested state}
\label{meta}
\end{center}
\end{figure}

Since for strong enough routing protocols (high $\eta$), in a certain range of $p$ there is coexistence
of the two phases, we would expect, because of finite size effects, that fluctuations
can trigger a jump from the free to the congested phase. As we can see in fig. \ref{meta}
this is the case for a square lattice of $100X100$ nodes, in which at $p=0.15$ $\eta=0.9$
the system, starting from the free phase, becomes congested after a while.
Moreover, once the system is congested into the coexistence region, the jump back into the free phase
is not possible anymore, because the congested phase is absorbing in the same way of the growing phase
of the pinning transition: the more the system remains in the congested phase, the more the queues
are growing and it is more hard to turn back into the free phase.
However, as the jump is triggered by independent activation processes,
the transient time goes exponentially with the systems' size.

\section{acknowledgments}
D.De Martino wants to acknowledge Grant No. PRIN 2007JHLPEZ (from MIUR), which partially supported this work,
D.Helbing, A.Vespignani and F.Caccioli for fruitful discussion.

\end{document}